\documentclass[letterpaper,useAMS,usenatbib]{mn2e}

\title[A large-scale CO survey of the Rosette Molecular Cloud]
{A large-scale CO survey of the Rosette Molecular Cloud: assessing the effects of O stars on surrounding molecular gas}

\author[W. R. F. Dent et al.]
{W.R.F. Dent$^{1,2}$\thanks{E-mail: wdent@alma.cl}, G. Hovey$^{3}$,
P.E. Dewdney$^{3}$, T. Burgess$^{3}$, A.G. Willis$^{3}$,
\newauthor
J.F. Lightfoot$^{1}$, T. Jenness$^{4}$, J. Leech$^{5}$, H.E. Matthews$^{3}$,
M. Heyer$^6$, and C. Poulton$^7$\\
$^{1}$UK Astronomy Technology Centre, Royal Observatory, 
Blackford Hill, Edinburgh EH9 3HJ, Scotland\\
$^{2}$ ALMA JAO, Av. El Golf 40 - Piso 18, Las Condes, Santiago, Chile\\
$^{3}$ Dominion Radio Astrophysical Observatory, PO Box 248, White Lake
       Road, Penticton, BC V2A 6K3, Canada\\
$^{4}$ Joint Astronomy Centre, 660 N. A'ohoku Place, Hilo, Hawaii 96720,
       USA\\
$^{5}$ Astrophysics Group, Dept. of Physics, Denys Wilkinson Building,
       Keble Road, Oxford, OX1 3RH, UK\\
$^{6}$ Dept. of Astronomy, Univ. of Massachusetts, Amherst, MA
       01003-9305, USA\\
$^{7}$ University of St. Andrews, Nth. Haugh, St. Andrews, Scotland \\
}

\begin{document}

\input{epsf.sty}

\date{}

\def\cm{\,{\rm cm}}
\def\cc{\,{\rm cm^{-3}}}
\def\asec{\,{\rm ''}}
\def\amin{\,{\rm '}}
\def\kps{\,{\rm {km~s^{-1}}}}
\def\Msun{\,{\rm M_{\odot}}}
\def\Lsun{\,{\rm L_{\odot}}}
\newcommand{\degree}{\mbox{\,$^\circ$}}        
\newcommand{\elec}{\mbox{\,e$^{-}$}}           
\def\elec{e$^{-}$}
%
\def\aa{{A\&A}}
\def\aar{{A\&ARev}}
\def\aas{{A\&ASup}}
\def\aj{{AJ}}
\def\apj{{ApJ}}
\def\apjl{{ApJLet}}
\def\apjs{{ApJSup}}
\def\araa{{ARAA}}
\def\ass{{A\&Sp.Sci.}}
\def\mnras{{MNRAS}}
\def\pasp{{PASP}}

\pagerange{\pageref{firstpage}--\pageref{lastpage}} \pubyear{}

\maketitle

\label{firstpage}

\begin{abstract}

We present a new large-scale survey of the J=3--2 $^{12}$CO emission
covering 4.8 square degrees around the Rosette Nebula. The results
reveal the complex dynamics of the molecular gas in this region. We
identify about 2000 compact gas clumps having a mass distribution
given by $dN/dM \sim M^{-1.8}$, with no dependence of the power law index
on distance from the central O stars. A detailed study of a number of the clumps
in the inner region show that most
exhibit velocity gradients in the range 1--3~$km s^{-1} pc^{-1}$,
generally directed away from the exciting nebula. The magnitude of the
velocity gradient decreases with distance from the central O stars, and we
compare the apparent clump acceleration with a photoionised gas
acceleration model. For most clumps outside the central nebula,
the model predicts lifetimes of a few $10^5 yrs$.
In one of the most extended of these clumps, however, a
near-constant velocity gradient can be measured over 1.7~pc, which is
difficult to
explain with radiatively-driven models of clump acceleration. 

As well as the individual accelerated clumps, an unresolved
limb-brightened rim lies at the interface between the central nebular
cavity and the Rosette Molecular Cloud. Extending over 4~pc along
the edge of the nebula,
this region is thought to be at earlier phase of disruption than
the accelerating compact globules.

Blue-shifted gas clumps
around the nebula are in all cases associated with dark absorbing
optical globules, indicating that this material lies in front of the
nebula and has been accelerated towards us. Red-shifted gas shows
little evidence of associated line-of-sight dark clouds, indicating
that the dominant bulk
molecular gas motion throughout the region
is expansion away from the O stars.  In addition, we find
evidence that many of the clumps lie in a molecular ring, having an
expansion velocity of 30~km$s^{-1}$ and radius 11~pc. The dynamical
timescale derived for this structure -- $\sim10^6 yrs$ -- is similar to
the age of the nebula as a whole ($2 \times 10^6 yrs$).

The J=3-2/1-0 $^{12}$CO line ratio in the clumps decreases with radial
distance from the exciting O stars, from 1.6 at $\sim$8~pc distance,
to 0.8 at 20pc. This can be explained by a gradient in the surface temperature
of the clumps with distance,
and we compare the results with a simple model of surface heating by the
central luminous stars.

We identify 7 high-velocity molecular flows in the region, with
a close correspondence between these flows and embedded young
clusters or known young luminous stars.
These flows are sufficiently energetic to drive gas
turbulence within each cluster, but fall short of the turbulent energy
of the whole GMC by two orders of magnitude.

We find 14 clear examples of association between an embedded young
star (as seen by Spitzer at 24$\mu m$) and a CO clump
in the molecular cloud facing the nebular.  The CO morphology
indicates that these are photo-evaporating circumstellar envelopes. 
CO clumps without evidence of embedded stars
tend to have lower gas velocity gradients.
It is suggested that the presence of the young star may extend the
lifespan of the externally-photoevaporating envelope.

\end{abstract}


\section{Introduction}

It is likely that a significant fraction of all solar-type stars
formed in molecular clouds
in the vicinity of massive luminous stars (eg Hester \& Desch, 2005).
An understanding of the interaction between OB stars and their molecular gas
environment is therefore necessary in order to understand star formation.
Such interaction has been the subject of considerable theoretical study (eg
Bertoldi \& McKee, 1990), and hydrodynamic models
show that complex dynamic structures are formed in such
environments (eg Lefloch \& Lazareff, 1994).  But the effects of OB
stars on
the formation of new lower-mass stars remains unclear:
both modest enhancements (eg Dale et
al., 2007) and supression of the star formation rate (eg McKee, 1989)
have been suggested.
Over the complete lifespan of a Giant Molecular Cloud however,
the presence of OB stars does limit the star formation efficiency,
due to the eventual dispersal of the molecular gas by ionising stellar
photons and winds (eg McKee \& Ostriker, 2007).
The interaction between OB stars and GMCs
have clear observational signatures, including ionisation fronts, Photon Dominated Regions (PDRs), bright-rimmed clouds, dark
clumps and cometary globules (eg Hester et al., 1996).  However,
optical and infrared observations of dark clouds and globules reveal only part
of the picture; they show neither the internal structure nor the
dynamics of the absorbing clouds, and they can only be seen
in the foreground of diffuse emission nebulae.

The Rosette Nebula, by dint of its overall symmetry and simple optical
morphology, has long been regarded as a textbook example of the
interaction between OB stars and surrounding gas.
The central cluster (NGC\,2244) contains six O stars with a total
luminosity of $\sim 10^6 L_{\odot}$ (Celnik 1985). Surrounding these is
the well-known symmetrical optical emission nebula.
To the southeast lies a large
molecular cloud complex, known as the Rosette Molecular Cloud
(hereafter ``RMC''), of total mass $10^5 M_{\odot}$, and extent
$\sim$40pc (eg Williams et al., 1995 - W95).  Near the centre of the
RMC lies the luminous embedded star AFGL961, well studied because of
its bright infrared spectra and outflow signatures (eg Lada \&
Gautier, 1982). The RMC itself has been mapped in the
J=1-0 transitions of $^{12}$CO and $^{13}$CO, with spatial resolutions
of 8~arcmin (Blitz \& Thaddeus, 1980), 1.7~arcmin (Blitz \& Stark,
1986) and 45~arcsec (Heyer et al., 2006). More detailed studies have
also been undertaken of individual objects within the RMC
(eg Patel et al., 1993; Schneider et al., 1998).

In the present work we describe new large-scale observations of the
Rosette Nebula and the surrounding region obtained with
high spatial resolution (14~arcsec) in the J=3-2 transition of $^{12}$CO.
This transition has an upper energy level of 33K and critical density of
$10^4 cm^{-3}$, as compared to 13K and
$300 cm^{-3}$ for the 1-0 transition, and so
are sensitive to relatively compact, warm clumps of
dense gas.  We use the data to take a census of
outflows within the RMC (section 3.1), look at the overall spatial and
velocity structure of the gas (3.2),
and investigate the compact clumps (3.3).
We compare the results with published CO~1--0 data
as well as observations of embedded young stars, in
order to determine the effects of the Rosette O-stars on the
surrounding GMC. We have adopted a distance of 1400~pc
for the Rosette Nebula (Hensberge et al., 2000).

\section{Observations}

The observations described in this paper were carried out using the
newly-commissioned heterodyne receiver array installed on the
James Clerk Maxwell Telescope, Mauna Kea, Hawaii.  The system consists
of HARP, the 16-element frontend receiver (Smith et al., 2003), and
ACSIS, the backend correlator and data reduction system (Hovey et al.,
2000). Controlling these instruments, as well as the telescope, is a
hardware realtime sequencer, as well as a high-level
software control system, known as the OCS (Rees et al., 2002).  For an
overall description, see Hills et al. (2008).

The large datacubes in the present observations were obtained
using the fast raster-scan observing mode of ACSIS/HARP.
The telescope was scanned at 75~arcsec/second
with a 10~Hz sampling rate, providing 7.5~arcsec sampling in the
scan direction.
The orientation of the regular square array of the HARP frontend was
maintained at $13\degree$ to the scan direction using a mechanical
beam rotator; with the diffraction-limited beam size of 14~arcsec and
the HARP beam separation of 30~arcsec, this resulted in near
Nyquist-sampled data. The chosen sampling was found to provide a good
compromise between wide area coverage and near Nyquist sampling of the sky.
The final image was constructed from several
map scans, typically $1\times0.5\degree$ or $1\times1\degree$ in
extent, each requiring 30-60 minutes to complete (including
pointing and observations of a spectral line
standard). In most of the mapped region, a basket-weaving technique was
employed, where the same area of sky was scanned in orthogonal
directions (RA and Decl., J2000). To improve the signal to noise ratio,
a second pair of orthogonal
scans were made in some regions of the map. During most of
the observations, two of the mixers were unusable 
and were flagged out during data reduction.
The basket-weaving observing technique significantly 
improved the fidelity of the final images in such circumstances,
by providing relatively
even sky coverage, better redundancy, and reduced scanning artifacts.
A common reference observation was taken every
60-100 seconds, at the end of every one or two rows. The line-free
reference position used was $06^h 35^m 10.0^s, +5\degree 23\arcmin 32.7\arcsec$
(J2000; these are the coordinates of the array centre).

A total of 20 hours of observations were taken in 2006 Dec~9 and 15
and 2007 Feb 14 and 15, including calibration and pointing. System
temperatures were mostly in the range 180-320K (SSB) during
the runs (although one scan was done at low elevation, resulting in system
temperatures as high as 450K). Main beam efficiency was determined to be
0.55$\pm0.05$, as measured on Jupiter (size 30~arcsec) and the JCMT
spectral line standard NGC2071IR. Measurements on the full Moon
showed maximum mixer-to-mixer calibration differences of about
5\% (Hills et al., 2009).
The pointing errors of the telescope at the time of these
observations were higher than normal, at 3~arcsec rms, as measured from
observations of the nearby source NGC2071. However, the largest
variation measured at the start and end of an individual raster map
was 7~arcsec, significantly less than the beam size (14~arcsec), and hence
these errors should not significantly affect the final map.

The $^{12}$CO J=3-2 line was observed in the lower sideband, at a rest
frequency of 345.796GHz.  ACSIS was set up in a configuration with a
total of 1.0~GHz bandwidth per mixer; with 2048 spectral channels per IF
and Natural weighting of the correlation function
this resulted in an effective spectral
resolution of 0.45$km s^{-1}$.  The data were reduced and calibrated
in real time using the ACSIS reduction software, which produces a time-
and coordinate-stamped series of spectra. The Starlink SMURF package
(Jenness et al., 2008) was used to transform these data into spectral
cubes, using a 7.5~arcsec grid spacing and assigning each data sample to its´ nearest neighbour in the final cube. The Starlink KAPPA and CCDPACK routines were used to remove
a linear baseline from each spectrum, transform to the same regular
coordinate grid, and combine the maps into a single output data
cube. To limit the final
datacube size, only the central 66 spectral channels (from $v_{lsr}=-2.9$ to  $25.1 km s^{-1}$) were included.

Individual spectra in the unsmoothed datacube had noise levels ranging from
0.9 - 3.6~K (T$_{mb}$, rms). Moreover, because of the missing mixers, a small number of pixels in this datacube were not filled.
To improve the fidelity of the images, the data were smoothed in the spatial
dimensions by a
Gaussian of width (fwhm) 15~arcsec, resulting in an effective image spatial
resolution of 20~arcsec. The spectral rms noise per pixel in
the final smoothed datacube then ranged from 0.2 - 0.5~K, with a mean value of 0.3~K (T$_{mb}$).
The final map, of size $\sim 2\times2.5\degree$, contained $10^6$ spectra.

\section{Results}

\epsfverbosetrue
\epsfxsize=14.0 cm
\begin{figure*}
\center{
\leavevmode
\epsfbox{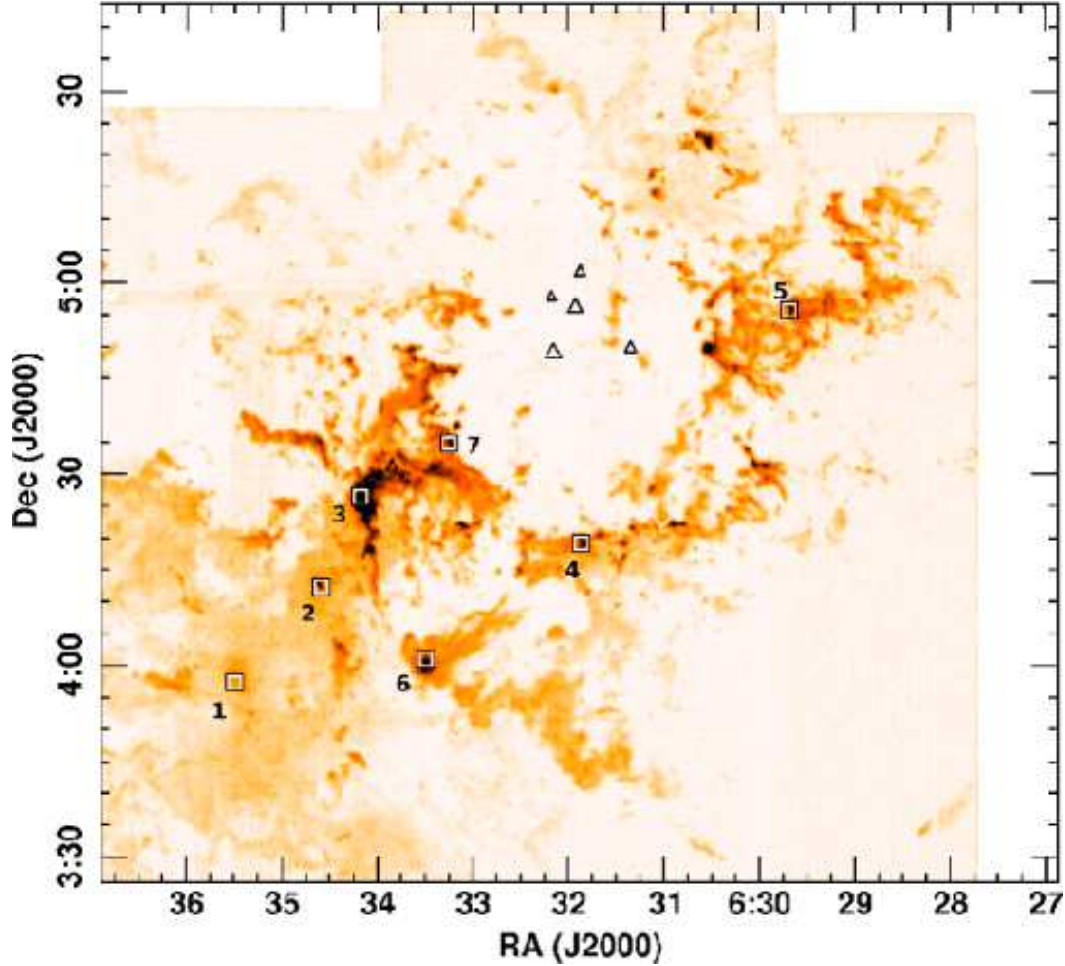}
\caption {Image of peak $^{12}$CO J=3-2 emission from the region of
the Rosette Nebula.  O stars illuminating the optical nebula are
shown as triangles, whose size represents their UV excitation
parameter (values from Celnik, 1985). Squares show the locations
and identification number of the outflows found
in the region (see section 3.1 and Table~1). The CO
greyscale ranges from 0 - 22 K (T$_{mb}$), and the image size
is 2.5x2.5 degrees (60x60~pc).}  }
\end{figure*}

An image of the peak CO J=3--2 emission, illustrating the overall structure of the
warm molecular gas in the region, is displayed in Figure~1. Also
shown are the locations of
the O stars illuminating the Rosette optical nebula; the size of the
symbols represents the UV excitation parameters of these stars (proportional to the radii of their Str\"omgren spheres - Celnik, 1985).
This figure illustrates the lack of molecular gas in the central region
around the most luminous O stars. The Rosette Molecular
Cloud, previously mapped at lower angular resolution in the CO~1--0
transition (Blitz \& Stark, 1986; Heyer et al., 2006) and (in part)
in the 3--2 transition (Schneider et al., 1998), appears as the
relatively bright complex structure to the SE of the O-star cluster,
extending over about $1\degree$ (20pc). In the area of overlap, the
new data agree reasonably well with the lower-resolution
large-scale images of Heyer et al. (2006). However, the factor of
$\sim3$ higher resolution and the higher energy level of the J=3--2
CO transition reveals more clearly the complex small-scale structures
in the warm gas throughout the cloud.
In particular, many compact clumps are clearly seen surrounding
the OB cluster and, by studying the full datacube, several
high-velocity outflows are found in the region (see
sections below). The {\em peak} brightness temperature shown in Fig.~1
emphasises the multiple compact (as well as some more
extended) clouds with narrow, bright emission lines.
The image also shows that
many of the clumps and clouds are limb-brightening on their edge
facing the O stars.

By consideration of the full datacube, we identify three distinct
spatio-velocity structures: (1) relatively compact regions with broad
line wings, identified as compact or well-collimated high-velocity
flows, frequently associated with bright IRAS sources; (2) bright,
compact clumps of emission, of sizes up to $\sim$3 arcmin ($\sim$1pc),
many of which have significant velocity gradients; and (3) extensive
undifferentiated regions of emission with narrow linewidths.  The
latter extended components were identified as ambient, relatively-undisturbed
cool gas clouds in the region (Heyer et al., 2006). In the
following sections we investigate the first two types of compact
structures in more detail.

\subsection{High-velocity outflows}

Inspection of the full datacube reveal seven potential
high-velocity (HV) protostellar outflows in the region; their locations and identifications
are marked by squares
on Figure~1. Such flows are identified in several ways. Firstly their
line profiles show broad wings of emission extending $\geq5 km
s^{-1}$ from the core cloud velocity, with a gradual decrease of intensity to higher relative velocities. This is substantially different in
character from the discrete, narrow velocity components that would arise from
clumps of ambient gas along the line-of-sight. As noted by Borkin et al. (2008),
inspection of full position-velocity datacubes can rapidly reveal these regions of HV gas in complex regions.
A confirmation of the HV outflow nature comes from the J=3--2 to J=1--0 line ratio in the wings. In all cases, this is significantly higher than the line core, implying that the HV gas is relatively warm.
Finally, the regions of HV gas are spatially relatively compact, no larger than a few arcminutes, which is unlikely for multiple line-of-sight clumps.

Table~1 gives the characteristics of these regions of HV gas, and
the CO J=3--2 spectral profiles at the central positions of the outflows
are shown in Figure~2. We also include in Fig.~2 the CO J=1--0 profiles for
comparison, using data taken by Heyer et al. (2006). The
J=3--2 data have been smoothed to the same spatial resolution as that of the
lower transition (45~arcsec) and both lines are shown on the main-beam
brightness temperature scale, T$_{mb}$.  In all cases, the line ratio in the wings lies in the range 1.5 -- 3, indicating excitation temperatures of 40-60~K, assuming
optically thin gas in LTE (cf Figure~12 below).
To illustrate the spatio-velocity structure in the outflows, Figures 3 and 4 show position-velocity cuts through the objects. These reveal the notable differences in characteristics of the ambient clumpy gas (often extended in the spatial direction, but compact in the spectral dimension), and the outflow gas (spatially-compact, but with high-velocity wings). Further searches through the datacube revealed no other regions with similar characteristics, although it is possible that some are hidden by multiple narrow line components along some lines of sight.

Blitz \& Stark (1986) proposed that broad CO emission from the RMC may be caused by high-velocity interclump gas. Conversely, Schneider et al. (1996) suggested that some of the apparent HV emission
is actually due to the superposition of multiple line-of-sight clumps. The higher spatial resolution of the present data supports their origin in HV outflows, at least for the seven objects noted above.
In Section 3.2, we discuss more the velocity extent of the compact clumps.

\epsfverbosetrue
\epsfxsize=12.0 cm
\begin{figure*}
\center{
\leavevmode
\epsfbox{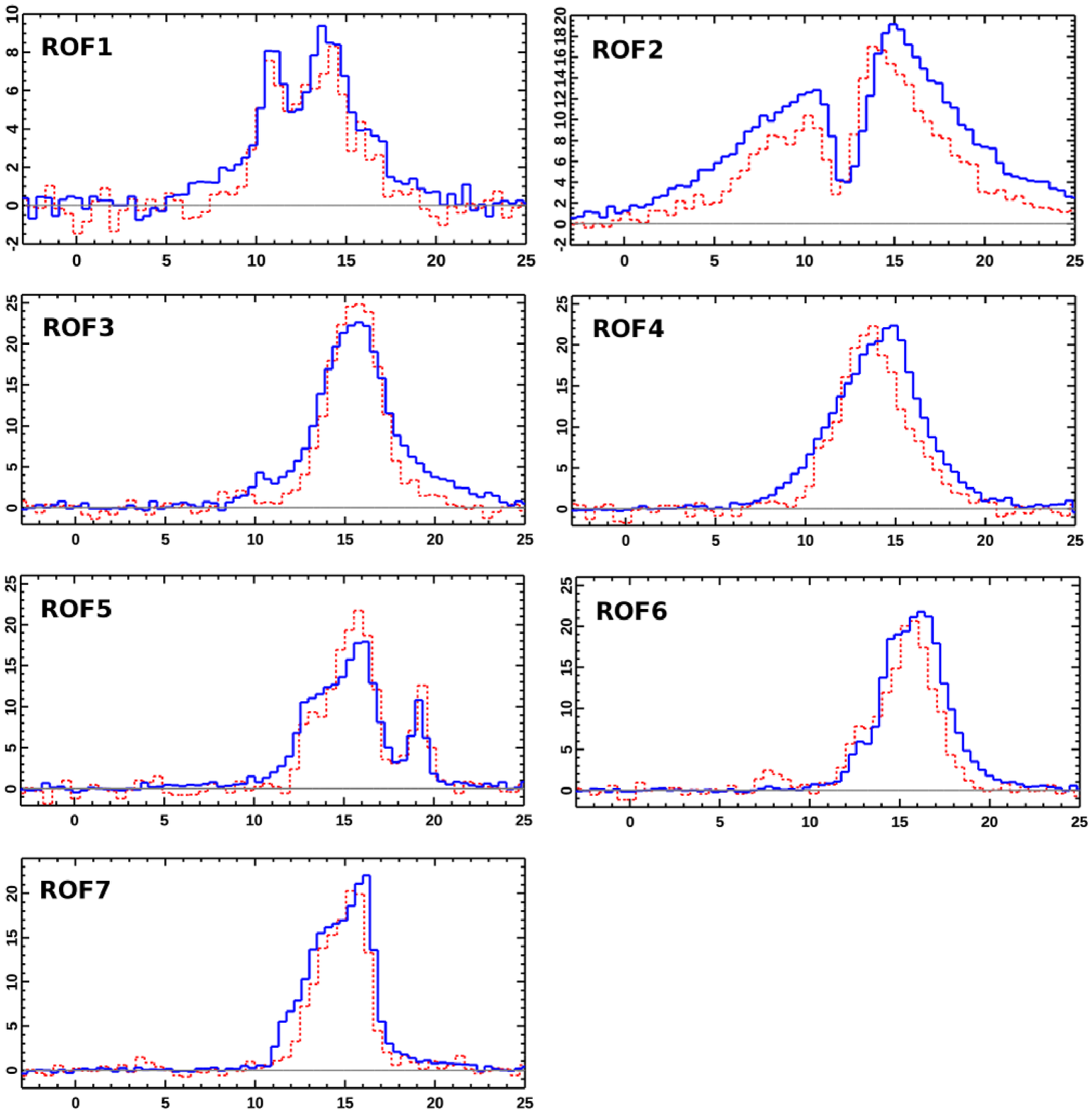}
\caption {Spectra at the centres of the high-velocity flows identified
in the region (ROF1-7); basic flow parameters are given in
Table~1. The solid blue histograms show the J=3-2 $^{12}$CO
line, and the dashed red histograms are J=1-0 data taken from Heyer et
al. (2005). The J=3-2 data has been smoothed to the same spatial
resolution as the lower transition (45~arcsec), and the intensities are
given on the T$_{mb}$ scale. Note that the spectrum of ROF2 extends beyond the
velocity region shown; these data were used to remove the DC baseline offset,
but were not included in the final datacube.}
}
\end{figure*}

\epsfverbosetrue
\epsfxsize=13.0 cm
\begin{figure*}
\center{
\leavevmode
\epsfbox{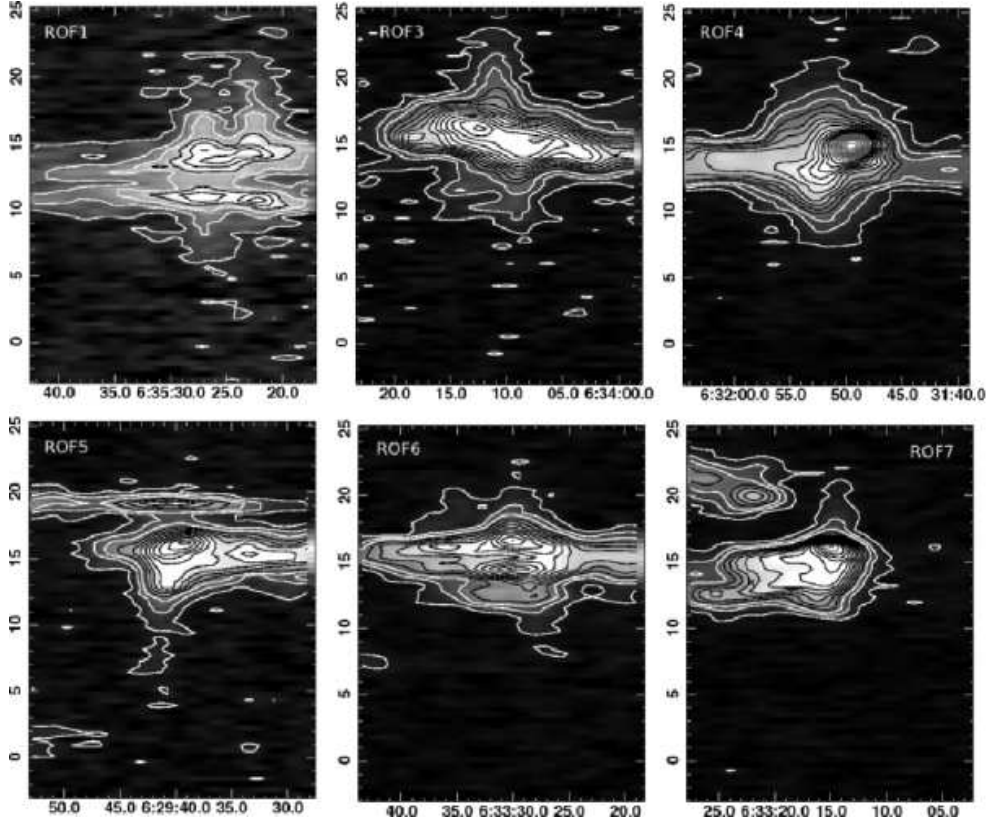}
\caption {Position-velocity cuts through the Rosette outflows ROF1 and ROF3-7. ROF2 has a more complex structure, and is shown in Figure~4. The x-axes are RA, and the y-axes are velocity (lsr). Lowest contours are 1~K, with a contour interval of 2~K
(T$_{mb}$). In all except ROF1, the centre of the cut corresponds to the coordinates in Table~1. For ROF1, the cut was through the peak HV emission to the SE of the centre (cf Figure~6).
}
}
\end{figure*}

\begin{table*}
\begin{center}
\caption{\sf Parameters of molecular outflows detected in the Rosette
field. See text for details. }
\begin{tabular}[t]{ccccclccccc} \hline
Name & RA & Dec & IRAS & Cluster & Outflow & Max. extent & Integ. flux
& Mass & KE$_{of}$ & KE$_{turb}$ \\
 & (J2000) & (J2000) & & & & (arcmin) & ($10^{33}K km s^{-1} m^2$) &
 (M$_{\odot}$) & ($10^{37}$~J) & ($10^{37}$~J) \\
\hline

ROF1 & 06:35:30.2 & 3:57:20 & 06329+0401 & PL7 & - & 4 & 17 & 3.4 & 22 & 764 \\
ROF2 & 06:34:35.2 & 4:12:19 & 06319+0415 & PL6 & GL961 & 5 & 71 & 14 & 275 & 791 \\
ROF3 & 06:34:10.7 & 4:26:29 & 06314+0427 & PL4 &HH871 & 2.5 & 21 & 4.0& 24 & 621 \\
ROF4 & 06:31:51.8 & 4:19:15 & 06291+0421 & PL1 & RMC-C,B(SII) & 2 & 10& 1.9 &11 &198 \\
ROF5 & 06:29:40.9 & 4:55:43 & 06270+0457 & - & - & 1.5 & 3.2 & 0.6 &2.0&-\\
ROF6 & 06:33:29.5 & 4:01:04 & 06308+0402 & PL3 & RMC-K(SII) & 2.5 & 6.3 &1.2&5.7&161 \\
ROF7 & 06:33:15.0 & 4:34:50 & 06306+0437 & PL2 & - & 1.0 & 2.4 & 0.5 &2.5&740\\

\\
 
\hline
\end{tabular}
\end{center}
\end{table*}

Table~1 gives the line intensities of the outflows
integrated over velocities beyond 
that of the ambient cloud emission. The velocity extent of the ambient
cloud was estimated by comparing the outflow profile
with the profiles of adjacent
outflow-free regions, estimated
using PV diagrams (see Fig~3, 4). Because of the ambient emission,
in most cases we cannot accurately measure
outflow emission within 3-5$km s^{-1}$ of the cloud velocity.
Outflow masses in Table~1 were obtained by integrating over the spatial
extent of the blue and red-shifted flows in the J=3--2 dataset, and assuming optically-thin
gas in LTE with T$_{ex}$=40K.  Both of these assumptions mean that the derived
masses of high-velocity gas are lower
limits; for a typical outflow inclination, this
method is thought to result in an underestimate of the total gas mass by a factor
of $\sim$3-10 (eg Beuther et al., 2002).  The outflow kinetic energy
(KE$_{of}$) was derived assuming a velocity equal to the maximum of
the line wing (typically 5--14~$km s^{-1}$).  Comparing the outflow
parameters in Table~1 with other published outflows, we find the
flows are typical of stars of comparable luminosity. For
example, taking the luminosities of the associated IRAS objects
(from Phelps \& Lada, 1997) and comparing these with the outflow
masses, we find that the all objects fall within the correlation of
outflows given by Wu et al. (2004). Hence
the Rosette outflows do not appear atypical.

\epsfverbosetrue
\epsfxsize=13.0 cm
\begin{figure*}
\center{
\leavevmode
\epsfbox{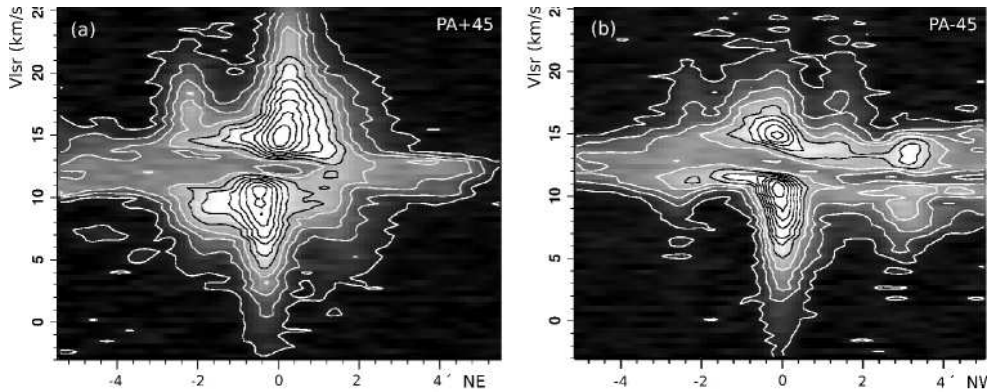}
\caption {Position-velocity plots of the Rosette outflow ROF2, taken (a) parallel to the main outflow (at PA=+45$\degree$, along the flow axis) and (b) orthogonal to the flow (at PA=-45$\degree$, but offset SW by 1~arcmin to show the bright HV lobes). The x-axis is the offset in arcmin (positive
offset in the left panel is NE along the main flow, in the right it is NW, orthogonal to the main flow); y-axis is the velocity (lsr). Lowest contours are 1~K, with a contour interval of 2~K (T$_{mb}$). Evidence of wings of HV emission, at velocities of $\pm 10 km s^{-1}$, can be seen along both axes, out to separations of $\pm$~2 arcmin from the central source.}
}
\end{figure*}

One object in Table~1 (ROF2) is a well-known outflow source
(AFGL961). Three others have published evidence of nearby shocked gas
from emission line surveys in the optical [SII] or infrared (H$_2$)
lines (Ybarra \& Phelps, 2004; Phelps \& Ybarra, 2005). The others are
new flows. Notably there is a close correspondence between this
list of
outflows and the list of embedded young clusters found by Phelps \&
Lada (1997): the only CO outflow not associated with a nearby cluster
(ROF5) lies outside their survey area (although
inspection of the 2MASS K-band images around ROF5 showed no bright
nearby cluster).  Conversely, of the 7 embedded clusters
identified by Phelps \& Lada, only one (PL5) has no apparent associated
outflow. Phelps \& Lada also identified 10 IRAS sources in the
region without an
associated cluster; none of these have evidence of outflows in
our survey.  Moreover, by inspection of the datacube in position-velocity space,
we find no evidence of HV gas in other regions
around the Rosette Nebula (see Figure~1). In summary, massive
high-velocity molecular outflows such as those listed in Table~1 are
closely associated with young clusters containing luminous IRAS objects.
IRAS objects without associated clusters do not show signs of HV outflows.
In the following sections we discuss some of these flows, first
individually, and then in the global context of the Rosette Nebula.

\subsubsection{AFGL961}

This was one of the first bright outflows to be discovered (eg Lada \&
Gautier, 1982), and published maps of the central region in
$^{12}$CO lines indicated that the flow direction lies
close to the plane of the sky (eg Schneider et al., 1998).

\epsfverbosetrue
\epsfxsize=12.5 cm
\begin{figure*}
\center{
\leavevmode
\epsfbox{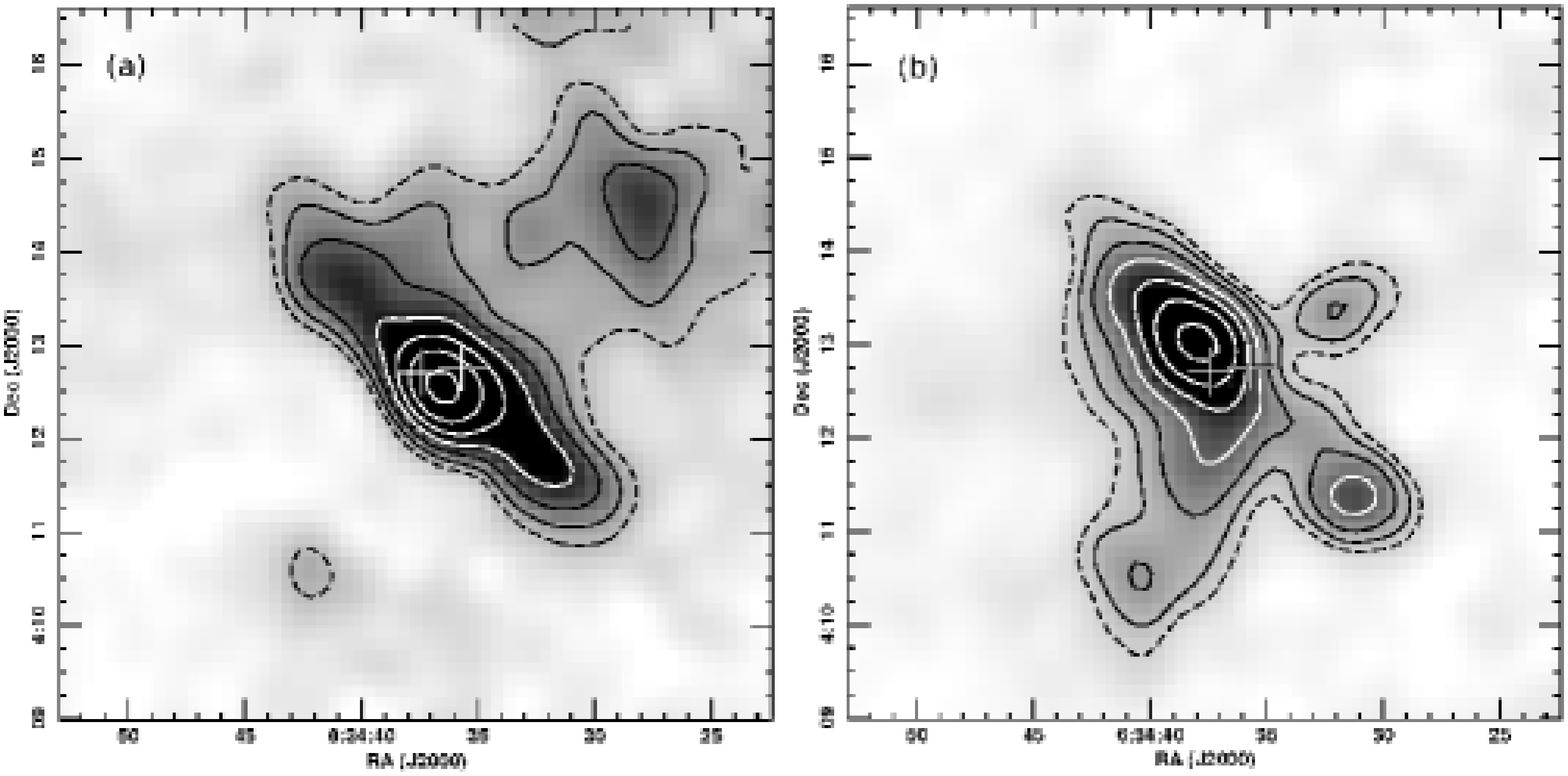}
\caption {Contours and greyscale images of integrated blue and
red-shifted gas in the region of the AFGL961 outflow (ROF2 in
Table~1). Velocity ranges
are: (a) -3 -- +10 and (b) +15 -- +25 $km s^{-1}$, respectively. The contours
start at $4K km s^{-1} (T_{mb})$ (2$\sigma$) and increment by factors of
1.5. The locations of the luminous infrared sources AFGL961 and AFGL961 II are shown by
crosses (using 2MASS coordinates; AFGL961 is the eastern object).}  }
\end{figure*}

Our new data (spectra in Fig.~2, and distribution of high-velocity gas
in Figure~4a and 5) show that the CO outflow profile taken towards the infrared source extends over $\sim30 km
s^{-1}$, and the main flow (at relative velocities up to $\pm 10 km s^{-1}$) is well-collimated in the NE-SW  direction, with a maximum extent of 5$\times$0.7 arcminutes
(2.0$\times$0.3~pc). The ends of the red and blue-shifted lobes are abrupt,
suggesting we are measuring the full spatial extent of the HV outflow. The
peak emission in the highest-velocity blue and red-shifted gas is separated by
$\sim$45~arcsec on either side of the bright IR source (AFGL961, the eastern of the two objects in Figure~5), indicating that this flow
is inclined to the plane of the sky. However, this inclination is
likely to be $<20\degree$, as the more extended red and blue-shifted lobes
both show similar morphologies. Comparison with infrared images
of H$_2$ v=1-0 S(1) emission from Aspin (1998) shows that the tip of
the NE lobe is concident with the most distant clump of shocked H$_2$ (CA6),
suggesting this is the bow shock at the end of the outflow jet.

Both the spectrum towards AFGL961 (Fig.~2) and position-velocity diagrams (Fig.~4) show a prominent
and narrow self-absorption dip near the RMC systemic velocity. This dip appears
spatially extended (over at least
$\sim$4 arcmin in Fig.~4), implying it is due to absorption in a large cool
cloud along the line-of-sight, rather than in a compact outer envelope around
AFGL961 itself.

In addition to the NE-SW flow, the integrated CO maps (Fig.~5) show high-velocity gas in regions extending orthogonal to the main flow axis. It can be seen as HV emission extending to $\pm$3~arcmin in the position-velocity cut of Fig.~4b.
If a separate flow, the origin appears to lie approximately 1~arcmin SW of AFGL961. This weaker
flow appears to be bipolar, with blue and red-shifted gas dominating
the NW and SE sides respectively, suggesting that the flow axis lies out of the plane of the sky. Infrared images of the region
show shocked gas is also found in several other directions in addition
to the dominant CO axis (Aspin, 1998).  AFGL961 itself is known to
contain several distinct components (eg Alvarez et al., 2004), and is
associated with a young cluster (Poulton et al., 2008). Moreover, recent infrared
images show the bright object 30~arcsec W of AFGL961 (known as AFGL961 II - see Fig.~5) has bipolar
shocked H$_2$ emission extending at PA$\sim 345\degree$ (Li \& Smith, 2005; Li et al., 2008). This may be associated with the second bipolar CO outflow.

In summary, the new extended high-resolution maps show that AFGL961 has a typical parsec-scale, well-collimated outflow, with evidence of a second orthogonal flow - possibly from AFGL961 II.

\subsubsection{Outflow ROF1}

\epsfverbosetrue
\epsfxsize=13.0 cm
\begin{figure*}
\center{
\leavevmode
\epsfbox{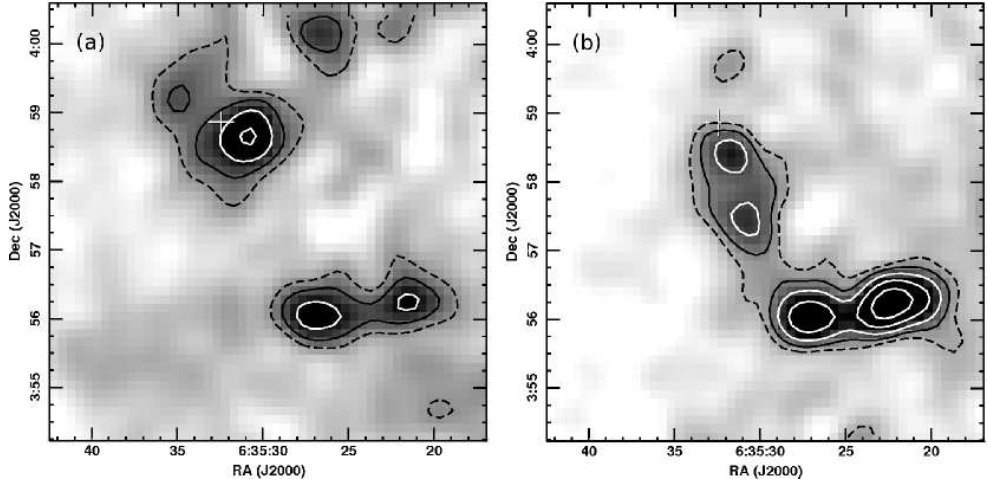}
\caption {Contours and greyscale images of integrated blue and
red-shifted gas from the ROF1 outflow. Velocity ranges are: (a) 6 -- 10 and (b)
15 -- 23 $km s^{-1}$. Contours start from $2K km s^{-1}$ and
increment by factors of 1.5. The location of the luminous infrared
source IRAS06329+0401 is shown by a cross.}}
\end{figure*}

The second region of clear spatially-extended HV wing emission is
ROF1; this is
closely associated with IRAS06329+0401 and cluster PL7 (see Phelps \&
Lada, 1997).  The distribution of HV gas is shown in
Figure~6. Both blue and red-shifted gas are found near the IRAS object, and there
are also more distant regions of HV gas at projected
separations of up to $\sim 1.6pc$. There is no embedded object near this
separate HV gas in the Spitzer survey of Poulton et al. (2008), so it
is unclear whether this is from a second fainter object or part of the flow
from the bright IRAS source. The coordinates given in Table~1 are mid-way
between the two regions.

\subsubsection{Ensemble outflow characteristics}

MacLow \& Klessen (2004) and others have suggested that gas turbulence within a GMC has a strong effect on the efficiency of star formation. Moreover, it has been proposed that protostellar outflows, such as those described above, could provide a significant contribution to this turbulence. Consequently it is important to assess the global effects of all outflows within a GMC.
The current dataset provides a census of HV outflows over a GMC of size
$\sim 60$~pc, which is complete down to an outflow mass of
$\sim 1 M_{\odot}$. This potentially allows us to assess the global effect of such flows on the RMC.
MacLow \& Klessen note that, in most cases, the sizes of outflows are generally too small to be important within a GMC.
The relatively compact nature of the outflows in the RMC confirms this.
Even the largest - AFGL961 - is only 2pc across, or less
than 5\% of the size of the RMC.  Within the individual young
clusters, however, the outflows have a similar extent to that of the
cluster itself, and their effect may be more significant.  We can estimate
the importance of the flows by comparing their kinetic
energy, (${KE}_{of}$ in Table~1) with the turbulent kinetic energy of the
associated large, dense molecular clumps (${KE}_{turb}$). Assuming the
linewidth $dv (km s^{-1})$ is from turbulent broadening, then
{KE}$_{turb} (J) \sim
5\times10^{35} M_c . dv^2$, where $M_c$ is the total
clump mass in units of $M_{\odot}$; values of $KE_{turb}$ are listed
in Table~1.  Using published clump masses
from the $^{13}$CO
J=1-0 observations of W95, the outflow kinetic energy
in Table~1 is 1-30\% of the clump turbulent energy. 
After allowing for an underestimate of the outflowing
mass by a factor of $\sim 3-10$ to account for inclinations and
optical depths (eg Beuther et al., 2002 - see above), then the outflow
kinetic energy is similar to the turbulent energy within the dense gas
clumps. On a larger scale, however, the main part of the RMC has mass and
velocity width of $7.5\times10^4$ and $\sim 4 km s^{-1}$ (W95), giving
${KE}_{turb}=10^{42} J$; this compares with the $1-3 \times 10^{40} J$
from all the outflows (after including the correction factor described
above).  Therefore, although the outflows may be energetically
significant on a stellar cluster scale, their effects are negligible on the
tens of pc scale of the whole RMC.

\subsection{Compact clumps and the inner edge of the cavity: 
global gas dynamics}

The molecular gas in the vicinity of the Rosette Nebula shows a
complex velocity structure, illustrated in the colour-coded image in
Figure~7. Here, we show integrated emission from gas which is
blue and red-shifted with respect to the RMC systemic
velocity, and (in green) the emission from gas around the ambient cloud velocity.  On the largest scale (tens of pc), the data shows a
velocity gradient from NW-SE of $\sim 0.07 km s^{-1} pc^{-1}$, similar
to that noted by W95. The distant gas is distributed relatively smoothly, both spatially and in velocity. However, this gradient is clearly highly
disrupted towards the central Rosette Nebula, where many compact clumps and
clouds at different radial velocities are superposed along the
line-of-sight.  Some distinct features are apparent in the
velocity-coded image which help in understanding the overall dynamics
of the system. Examples are the NW ridge - a line of blue-shifted
clumps seen most clearly $\sim10$~pc NW of the centre (along the NW of the ellipse in Fig.~7), and the Monoceros Ridge - the
abrupt edge of the SE molecular cloud $\sim10$~pc SE of the centre (seen along the SE of the ellipse). These distinct features are described in the following sections.

\epsfverbosetrue
\epsfxsize=12.0 cm
\begin{figure*}
\center{
\leavevmode
\epsfbox{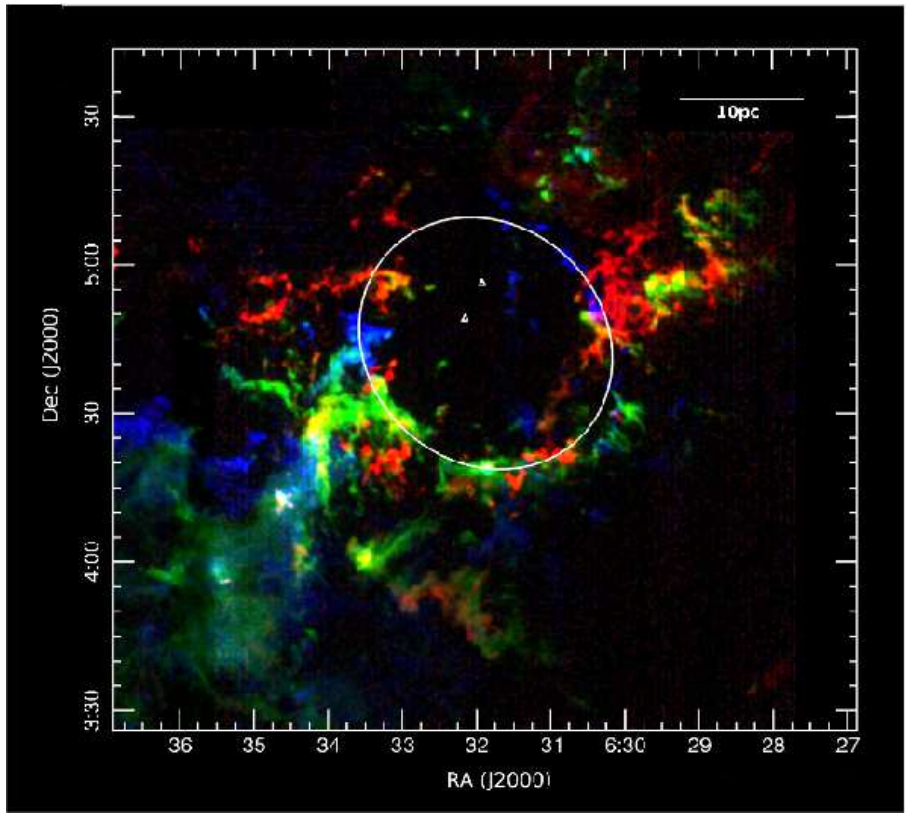}
\caption {Colour-coded image of the molecular cloud around the Rosette
  Nebula, illustrating the global velocity structure.  The blue, green
  and red colours represent integrated $^{12}$CO intensities in the
  velocity ranges (-2.8,+11.5), (+11.5,+16.6) and (+16.6,+25.1) $km
  s^{-1}$. The small triangles near the centre show the locations of the
  two most luminous O stars in the nebula (HD46223 and HD46150), and
  the ellipse shows a model of an inclined molecular ring, of radius $\sim11$pc,
  and inclination 30$^o$ to the line-of-sight (see text for
  details). The image is 61~pc (EW) $\times$ 56~pc (NS), assuming a
  distance of 1400pc.}

}
\end{figure*}

The NW ridge is a line of compact blue-shifted clumps to the NW of the
central O stars. These molecular clumps are all
associated with dark absorbing globules and Elephant Trunks
in the optical images (see section 3.3, Figure~9). Their high optical
obscuration and blue-shifted profiles indicates they are on the
nearside of the nebula, moving towards us (with respect to the
systemic cloud velocity). Moreover there is a clear velocity gradient
along the ridge; evidence of this can be seen in the colour-coded
image
in Figure~9. In order to help understand these data, we have
fitted the positions, velocities and velocity gradients of the
emission clumps using simple 3D Gaussians (see section 3.3.2). Using
the centroid velocities of these fitted clumps, we investigate the
velocity structure along the inner rim of molecular gas around the
Rosette Nebula (to a projected radius of $\sim 15pc$ from the
centre of the nebula, and a depth of 3.5pc (9 arcmin) from the inner
edge of the rim). Figure~8 shows the velocity of these clumps as a
function of $\phi$, the azimuthal angle around the rim. The velocity
structure of much of the NW blue-shifted ridge (at $0 \leq \phi \leq
120^o$) can be most easily
fit with that of an expanding ring of radius 11pc,
inclined at 30$^o$, with an expansion velocity of 30$km s^{-1}$ and a
mean velocity of $16 km s^{-1}$.  This is illustrated by the ellipse
in Figure~7. Evidence of gas clumps fitting such a ring can be seen in
regions of $\phi$ up to 180\degree.
Around the SE rim ($-180 \leq \phi \leq -90^o$) there
are few corresponding red-shifted clumps as predicted by the simple
expanding ring
model. Most emission here lies within $3km s^{-1}$ of the RMC
systemic velocity, suggesting this material is not participating in
the ring expansion.  Nevertheless, the structure and velocity
gradients across the SE cloud rim clearly indicate that this cloud is
exposed to the O star radiation field; this suggests that expansion on
this side may have been confined by the ambient cloud. The conditions
of this cloud rim will be described further in section 3.3.2.  Some
additional
blue-shifted emission is apparent on the E rim, at $\phi\sim-90^o$ in
Figure~7 and 8, and this corresponds to an optically-dark obscuring cloud
(eg Gahm et al., 2007). Notably, {\em all} of the dark
clouds in the region of the Rosette Nebula
- which are presumed to be in the foreground of the nebula -
appear to be blue-shifted with respect to the cloud systemic
velocity. This strongly suggests that all these
foreground clouds are expanding away from the nebula towards us, and
that most (but not all)
of this expansion is in the form of a ring of material.

\epsfverbosetrue
\epsfxsize=8.0 cm
\begin{figure}
\center{
\leavevmode
\epsfbox{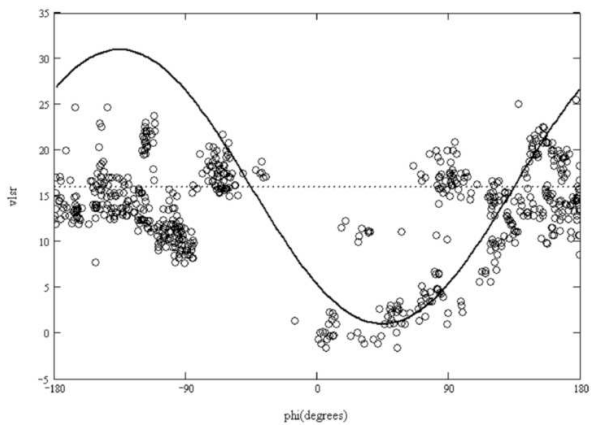}
\caption {Radial velocity of CO clumps around inner rim of the Rosette
  Nebula plotted as a function of $\phi$, the position angle.  The
  solid line illustrates the model of an inclined (i=30$^o$),
  expanding ($V_{exp}=30 km s^{-1}$) ring of radius 11pc, with a mean
  velocity of $16 km s^{-1}$ (given by the dotted line).  See text for
  details.}

}
\end{figure}

Radio recombination lines show the ionised gas in the nebula has a
mean velocity of 16.7$km s^{-1}$ (close to that of most of the CO
emission - see Fig.~8) and a line fwhm of 30-40 $km s^{-1}$
(Celnik, 1985). This is consistent with the expansion velocity of the
clumpy ring.  Assuming a constant acceleration, then the expansion
velocity and ring size implies a dynamical time of $\sim0.8$~Myr,
comparable with estimates of the system age of 2-3~Myr (Hensberge et
al., 2000; Balog et al., 2007).  The Rosette expanding ring
has characteristics similar to the large molecular structure
found around the nebula IC1396 (Patel et al., 1995). Here they have found
a ring of clumpy gas with an expansion velocity of $5 km s^{-1}$ and
radius 12pc, surrounding an O6 star of age 2-3Myr.

\subsection{Clumps in and around the Rosette Nebula}

It has been known for many years that the Rosette Nebula is rich in
dark obscuring material, first seen in optical images (eg Herbig, 1974).
These structures range from extended dark globules and Elephant Trunks many arcminutes in extent (eg Gahm et al., 2006), down to
``Teardrops'' (eg Herbig, 1974) and ``Globulettes'' on scales of
arcseconds or less (Gahm et al., 2007).  Clearly, however, optical observations
are limited to the dark absorbing regions in front of the diffuse nebular
emission.  Observations of the
molecular gas, which have the advantage of tracing material thoughout
the region, also show a highly clumped structure, albeit previously
limited by
observations to arcminute, or larger, scales (W95).
Many of the largest optically-dark regions, however, have
corresponding CO emission clumps (eg Schneps et al., 1980).

In the NW ridge, the present high-resolution data show that
{\em all} of the blue-shifted CO clumps have associated obscuring
optical regions, implying there is no blue-shifted gas lying behind
the nebula: Figure~9 compares the distribution of blue-shifted gas
(at velocities $v_{lsr}< 4.7 km s^{-1}$)
with the optical H$\alpha$ image from the IPHAS survey (Drew et al., 2005).
Conversely, {\em none} of the optically-dark clouds are
associated with red-shifted gas, implying there is no foreground
gas receding from us. An additional region of blue-shifted gas
$\sim$20pc SE of the O stars can be seen in Figure~7. This also
corresponds to obscuring foreground clumps (Patel, 1993), confirming that
the dominant molecular gas motion in the RMC is expansion towards us, away
from the nebular region.

Focussing on the NW ridge, in Figure~9 we have labelled some of the most
prominent CO clumps facing the nebula, and in Table~2 we list
their main parameters (clumps 1-15).  Gahm et al. (2007) conducted optical
studies of the highly compact globules in this region, and comparison with
our new observations indicates that the present survey is able
to detect CO emission from
isolated globules with masses as small as $\sim 0.03-0.1
M_{\odot}$. Examples of these are the three faint clumps
1~arcmin S of the cloud marked 6 in Figure~9, which correspond to
Globules 47, 40 and 35 in Gahm et al. (2007). However, most clumps
in the datacube
lie in more complex and confused regions, which means the lower limit to
their detectable mass is larger than this (see section 3.3.3).  In the
following sections, we discuss some of the salient
features of the individual clumps.

\epsfverbosetrue
\epsfxsize=16.0 cm
\begin{figure*}
\center{
\leavevmode
\epsfbox{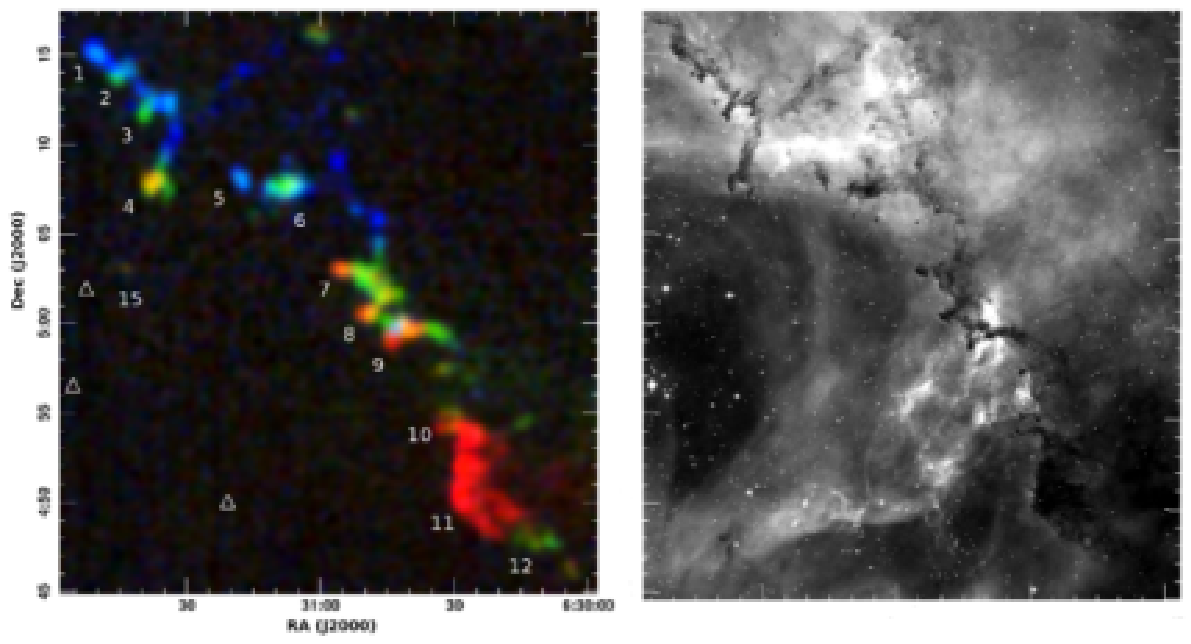}
\caption {Colour-coded image of the blue-shifted CO emission in the NW
ridge (left), compared with an optical H$\alpha$ image of the same
region (image from the IPHAS survey, right). The exciting O stars of
the nebula lie to the SW of these panels.  The CO map is shows peak
intensity, taken over 3 different velocity regions represented by
different colours, where blue is $-2.8,-0.3 km s^{-1}$, green is
$-0.3,+2.2$ and red $+2.2,+4.7 km s^{-1}$; this compares with the
systemic cloud velocity of $+16 km s^{-1}$.  All of the molecular
clumps can be identified with dark clouds in the optical image,
implying all lie in front of the nebula. The main CO clumps are
labelled, and their parameters are given in Table~2.  All clumps,
aside from \#12 and 14, have a velocity gradient with increasing blue
shift as a function of projected separation from the O~stars (marked by the white triangles).}  }
\end{figure*}

\subsubsection{Clump parameters: dynamics and physical conditions}

The colour-coded image in Figure~9 illustrates the detailed
velocity structure of the NW clumps.
Significant velocity shifts were known to exist across three
of the brightest extended clumps in this region (Schneps et al., 1980; Gahm
et al., 2006).  The clearest example of a velocity gradient is seen in the
``Wrench'' trunk (marked as 4 in Figure~9).
The resolution of the new dataset is sufficient to show that
the majority of these NW clumps have radial
velocity gradients of $\delta v/\delta r >1 km s^{-1} pc^{-1}$; these are
listed in the upper section of Table~2. In most of the well-defined
NW clumps, the highest gradient occurs radially
from the central O stars, in the sense that the blue-shift
increases with increasing separation ($\delta v/\delta r < 0$).
For comparison with the NW clumps, the middle sections of
Table~2 list
parameters of several other distinct structures in the Rosette
region, including bright inner clumps, bright regions
of the sharp-edged rim at the
interface between the nebula and RMC,
and more distant and isolated clumps in the surrounding cloud. 
Finally, the lowest section in Table~2 lists parameters of
centres of some of the smooth extended clouds seen in the datacube.
These diffuse clouds were studied by Heyer et
al. (2006) and are thought to be diffuse ambient molecular gas
undisturbed by the activity of the O stars in the Rosette Nebula.

\begin{table*}
\begin{center}
\caption{\sf CO line parameters of clumps and distinct objects in the vicinity
of the Rosette Nebula.}
\vskip -0.8cm
\begin{tabular}[t]{lccccccclrcl} \hline\

Clump &RA&Dec&v$_{lsr}$& $\delta$v/$\delta$r$^{(a)}$ & $\Delta$v$^{(b)}$ & $T_{mb}$3-2$^{(c)}$ & $T_{mb}$1-0&Ratio$^{(d)}$&D$^{(e)}$&Star&Notes\\
&(J2000)&(J2000)&($km s^{-1}$)&($km s^{-1} pc^{-1}$) & ($km s^{-1}$) & (K) & (K) & &(pc)&$^{(f)}$&\\
\hline
\multicolumn{4}{l}{\it{NW clumps}}\\
1   & 6 31 50.2&  5 15 00&-0.6& -0.9 &1.4 &   10.2  &  7.8 &   1.3&10.6&o&\\   
2   & 6 31 46.2&  5 13 50&0.4& -2.2  &1.9  &  8.0  &  8.9 &   0.9  & 10.2 &o&\\  
3   & 6 31 39.4&  5 11 47&0.8& -1.9  &1.8  &  7.8  &  5.5 &   1.4  & 9.6&*& \\ 
4   & 6 31 36.9&  5 07 52&0.3& -3.7   &2.0  &  11.8  &  9.8 &   1.2  &8.2 &*&Wrench clump$^{(1)}$\\ 
5   & 6 31 18.0&  5 07 55&-0.1& 1.1 &1.7  &  6.7  &  7.6 &   0.89  &  9.2  &o&\\ 
6   & 6 31 08.3&  5 07 34&0.7& 1.0 &2.1  &  10.7  &  11.1 &  0.97 &9.7&o&Clumps superimposed?\\  
7   & 6 30 54.8&  5 02 57&2.4& -2.0 &1.3  &  10.2  &  8.9 &   1.2  & 9.4 &-& \\  
8   & 6 30 49.5&  5 00 26&2.5& -1.7 &1.8  &  8.9  &  8.9 &   0.9  & 9.3 &-& \\  
9   & 6 30 44.7&  4 58 57&2.7& -1.6 &1.4  &  6.6  &  7.1 &   0.92  & 9.5  &-& \\  
10  & 6 30 32.8&  4 57 24&1.9& -1.9 &1.2  &  3.7 &  3.6 &   1.0  & 10.4&-& \\  
11  & 6 30 28.7&  4 54 11&3.2& 2.3 &1.4  &  8.7  &  8.7 &   1.0  & 10.5&-& \\  
12  & 6 30 26.5& 4 50 13&5.1& 1.6 &2.2  &  14.2  &  18.9 &   0.75  &10.5&-&Unclear\\   
13  & 6 30 11.1&  4 47 42&1.4& -1.8 & 1.3 &   4.0 &   3.6 &   1.1 & 12.1&-& \\   
14  & 6 30 00.1&  4 43 00&2.7& -0.9 &2.0  &  5.1  &  3.6 &   1.4 &13.5&-&\\
15  & 6 31 43.7&  5 03 06&2.2& 0.2  &2.4  &  1.7  &  1.2 &  1.5  & 6.1&*&\\ 
\multicolumn{4}{l}{\it{Other inner clumps \& rim}}\\
16  & 6 33 15.8&  4 30 37&15.0 &  2.0 &2.8  &  16.6  &  20.1 &   0.8  &10.2&*&Monoc. Ridge$^{(2)}$\\
17  & 6 32 29.1&  4 39 57&14.2& 1.6 &1.6  &  10.2 &  11.6 &   0.89  & 4.4&o& \\
18  & 6 33 14.1&  4 46 17&10.3& -2.6 &2.1  &  16.6 &  18.4 &   0.9   & 6.6&*&Extended Ridge$^{(2)}$\\  
19  & 6 30 30.7 &  4 49 27&15.9&2.1 &1.6  &  21.1 &  22.4 &  0.94 & 10.1&-& \\   
20  & 6 32 39.2&  4 55 19&15.6& 2.7 &1.7  &  9.8 &  10.9 &  0.9  & 3.7&o&  \\  
21  & 6 32 29.3&  4 30 32&15.9&  -2.4  &1.1  & 10.7 & 6.7 &  1.6  &8.0&o& \\   
22  & 6 32 39.4&  4 26 37&13.0&  2.4 &1.5  &  12.6  &  10.7 &  1.2 & 9.8&o&SSW rim\\
23  & 6 31 07.6&  4 27 37&16.5&  0.7 &0.8   & 16.2  &  23.3 &   0.69  & 11.0&*& \\
24  & 6 33 10.7&  4 37 37&20.6&  0.9 &1.9  &  20.3 &  16.7 &   1.2  & 7.8&o&   \\
25  & 6 32 01.0&  4 27 28&13.7& 0.7 &1.5  &  6.2 &  5.6 &  1.1  & 9.1&o& \\  
26  & 6 32 59.5&  4 30 15&14.3&  2.8 &1.4  &  14.0  &  13.3 &   1.0  &9.4&*&Monoc. Ridge$^{(2)}$\\  
27  & 6 31 02.8&  4 20 27&13.1&  0.2&1.4  &  8.7  &  10.0 &   0.87  &13.7&-&S rim\\  
28  & 6 31 25.2&  4 19 08&14.4&  0.7 &1.0  &  17.1  &  14.7 &   1.2 &13.2&o& S rim\\  
29  & 6 33 08.4&  4 47 01&10.9&  -1.3 &1.5  &  7.6  &  6.7 &   1.1 &  6.0&*&Pillar$^{(3)}$\\
30  & 6 31 51.2 & 4 23 13&15.5&  -0.5 &1.1  &  6.6  & 6.0 & 1.1 & 10.9 &o&\\
31  & 6 32 59.2 & 4 23 44 &17.2&  2.7 &1.9  &  10.7  & 9.8 & 1.1 & 11.6 &*&\\
32  & 6 33 22.6 & 4 40 44 &10.4&  1.4 &1.6  &  13.3  & 8.4 & 1.6 &  8.2 &*&Monoc. Ridge$^{(2)}$\\
33  & 6 31 20.6 & 4 22 55 &14.9&  -2.3 & 1.5 &  5.8 & 6.2 & 0.94 &12.0&*&S rim\\
34  & 6 33 08.0 & 4 40 13 &17.0&  1.5 & 2.7 &  10.2 & 10.2 & 1.0 & 7.0&o& \\
\multicolumn{4}{l}{\it{Distant clumps}}\\
35  & 6 32 08.9&  4 05 15&14.2&-0.2&  0.7 &  8.7  &  10.7 &   0.82   & 18.1&o& \\  
36  & 6 33 21.5&  4 19 05&18.2& 1.7 &2.0   &  11.8  &  10.7 &  1.1  & 14.4&o&\\   
37  & 6 34 18.6&  4 47 42&11.2&  -0.8 &1.6  &  15.8  &  15.6 &   1.0  & 13.0&o& \\
38  & 6 34 14.1&  3 45 40&12.7&  -0.5 &0.9  &   12.2 &   12.9 &    0.95  & 28.9&o& \\   
39  & 6 30 36.9&  5 23 55&12.8&  1.3 & 1.4 &  18.7 &  22.0  &  0.85  & 16.6&-&  \\  
40  & 6 28 32.2&  4 58 58&15.8&  -0.6 &1.3  & 12.7  & 15.6  &  0.82   & 22.5&-& \\  
41  & 6 29 55.5&  4 31 16&14.4&  1.1 &2.0  &  19.1 &  18.2  &  1.0  & 15.6&-& \\  
42  & 6 34 24.0&  3 48 41&14.7&  -0.1 &0.5  &  8.2  &  8.9  &  0.92  & 28.3&o& \\ 
43  & 6 30 34.0&  4 10 07&14.2&  -0.4 &0.9  &  4.6  &  4.7  &  0.97  & 18.9&-& \\  
44  & 6 35 02.0&  4 25 19&8.2&  -1.7 & 1.1  &  12.9  &  14.4  &  0.89  & 20.2&*&IRAS source$^{(4,5)}$\\  
45  & 6 33 54.9 & 4 04 50&10.1&  0.6 & 0.5  & 2.6    & 3.1    & 0.83 & 21.1&o&\\
46  & 6 34 01.2 & 4 02 23&9.6&  -0.3 &0.8  & 4.2 & 3.9 & 1.1 & 22.3 &o&\\
47  & 6 35 01.0 & 4 21 17&13.7& -1.7 &0.8  & 7.6 & 9.1 & 0.84 & 20.8&*&\\
48  & 6 32 06.4 & 4 15 10&16.7&  -0.6 &0.7  & 12.2 & 14.7 & 0.83 & 14.0 &*&\\
49  & 6 31 48.2 & 4 11 34&17.3&  0.8 &0.9  & 10.5 & 7.8 & 1.4 & 15.6 &o&\\
\multicolumn{4}{l}{\it{Extended smooth clouds}}\\
50  & 6 35 46.0&   3 39 10&11.1&  -0.4 &1.6  &  4.2  &  4.9  &  0.86   & 36.1&o&\\  
51  & 6 30 10.0&   5 19 00&17.8&  0.5 &1.0  &  2.9 &  7.1  & 0.41  & 17.1&-&\\
52  & 6 31 55.7&  3 44 09&16.2&  -0.4 &1.3  &  7.4  &  10.2  &  0.73  & 26.7&o& \\  
53  & 6 32 47.5 &  5 36 56&15.9&  -0.5 &0.4   &  4.0 &  6.8 &  0.59  & 19.7&-&  \\  
54  & 6 36 31.8&  4 27 04&5.1&  0.1 &1.4  &  2.7 &  5.5 &  0.5  & 28.0&o&  \\ 
55  & 6 28 10.2&  5 00 46&19.8&0.1 &0.9  &  2.0 &  5.6  &  0.36  & 24.4&o& \\ 
56  & 6 34 01.8&  5 18 10&10.8&  0.3 &0.5  &  4.6 &  6.0  &  0.76  & 16.2&-&  \\ 
57  & 6 25 57.6&  4 09 50&14.8&  0.3 &1.1  &  5.3  &  8.7  &  0.61  & 28.1&-&  \\ 

\hline
\end{tabular}
\end{center}

\newpage

Notes to Table~2: (a) Velocity gradient over 60--100 arcsec
distance around given coordinate, measured radially from most
luminous O star (HD467223); (b) Velocity width (fwhm) of J=3-2
$^{12}$CO line from full spatial resolution map, after deconvolution
with instrumental velocity resolution ($0.45 km s^{-1}$); (c) Main
beam brightness temperature of $^{12}$CO J=3-2 line, after smoothing to
same spatial resolution
as J=1-0 data (45~arcsec); (d) Ratio of $T_{mb}$; (e) Projected separation
from HD46223; (f) An asterix indicates a star or compact infrared
source (from Spitzer 8$\mu m$ and/or 24$\mu m$ images)
can be identified with the molecular clump, 'o' indicates no star, '-'
indicates region not covered by Spitzer images.
References: (1) Gahm et al. (2006); (2) Schneider et al. (1998);
(3) Balog et al. (2007);
(4) Patel et al. (1993); (5) White et al. (1997)

\end{table*}

On the SE side of the central cavity, most of the clumps also have
significant velocity gradients, and most have $\delta v/\delta r > 0$.
Optical images show that these molecular clumps do not have clear
corresponding dark clouds, and therefore lie towards the far side of the
nebula. This gas appears to be accelerating away from us on the far side.
There is one notable exception to this: the red-shifted clump 8~pc SE of the
central star (Clump 21, see Fig.~7), which appears to be
accelerating in the opposite direction. Inspection of the
optical image shows this actually corresponds with a dark obscuring cloud, so
it lies in front of the nebula and is an isolated clump
undergoing acceleration towards us.

Assuming the velocity gradients are due to radial motion away from
the central O stars, then the radial clump acceleration is given by:
$$ a_c = 0.5 [(v_c.{\delta~\!v} + {\delta~\!v}^2)/ {\delta~\!r}]$$
Here $v_c$ is the clump velocity relative to the molecular cloud
systemic velocity and
$\delta~\!v$ the velocity difference along the cloud of length $\delta~\!r$.
The apparent clump acceleration, $a'$, depends on the inclination of the
accelerating surface to the plane of the sky: $a' = a_c /({sin}^2~\!i.cos~i)$.
In Figure~10 we show the radial acceleration of the clumps in Table~2
assuming $i=30\degree$ (equal to that of the expanding ring in section 3.2),
plotted as a function of projected distance from the centre of the nebula.
The RMC systemic velocity of $16 km s^{-1}$ was used to determine $v_c$.
The plot shows a general decrease with increasing separation,
suggesting that clump acceleration is caused by the stars in the
centre of the nebula. However, some clumps have relatively low
apparent velocity gradients; these may be moving closer to the plane of the sky
than the inclined ring.

Observations of molecular gas clumps and globules near other
nebulae show velocity gradients of a few
$km s^{-1} pc^{-1}$, similar to those in Table~2 (eg Lefloch \& Lazareff,
1995; Gonzalez-Alfonso et al., 1995; Pound, 1998; Bachiller et al., 2002).
One model for the acceleration of molecular gas in such clumps
is through a rocket effect, whereby the clumps are
gradually photoevaporated by the O-star photons,
and the pressure from the photoionised wind from
the ablating surface drives the clump away from the star (eg
Oort \& Spitzer, 1965; Bertoldi \& McKee, 1990).
Bertoldi \& McKee (1990) have developed this
Radiatively-Driven Implosion (RDI) model and derive (in their eqn.~4.2),
the resulting mass loss rate of clumps of radius
$R_{c,p} (pc)$ and mean density $n_{H_2} (cm^{-3})$. From the mass loss rate,
we can estimate
the resulting clump acceleration:
$$a = 5.6\times 10^{-6} . \phi . cos~{i} . S_{_{50}}^{0.5}. D_p^{-1}.
R_{c,p}^{-1.5}. V_{r,_{10}}. n_{H_2}^{-1} \ \  [{m}{s}^{-2}]$$
where $D_p$ is the projected separation
from the O star (in pc), $S_{_{50}}$ the
ionising flux from the star(s) in units of $10^{50} s^{-1}$
(equal to the estimated total ionising flux from the Rosette O stars),
and $i$ is the projection angle of the clump from the plane of the sky.
The Rocket velocity of the ionised particles is
$V_{r,_{10}}$ (in units of $10 km s^{-1}$ - typical for most
models), and $\phi$
is a dimensionless factor governing the clump mass loss rate, taken to be
$\sim 3.0$ for the conditions in the Rosette Nebula (see Bertoldi \&
McKee, Fig.~11).
The solid line in Figure~10 shows the predicted clump acceleration as
a function of distance from the O stars, for
clumps of radius 0.1~pc (30~arcsec diameter) and density $3\times 10^4 cm^{-3}$
(giving a clump mass of 8$M_{\odot}$ - similar to that of the Wrench Clump
measured by Gahm et al., 2006). The model fits the upper envelope
of the data points, using an inclination of $30\degree$;
the Wrench clump and other NW clumps have the highest acceleration in
the plot and fall near the model line. In general, however, most clumps lie
below the model prediction, suggesting they have a
lower inclination (lying closer to the plane of the sky),
or that the acceleration efficiency of distant clumps
is lower than predicted.

Also shown in Figure~10 is the estimated lifetime to photoevaporation.
For clumps within 5pc of the O stars, this is very short - less than
10\% of the age of the Rosette Nebula. Even the distant clumps have
relatively short lifespans compared with the nebula age of $\sim 2 Myr$,
implying they are transient phenomena.

\epsfverbosetrue
\epsfxsize=9.0 cm
\begin{figure}
\center{
\leavevmode
\epsfbox{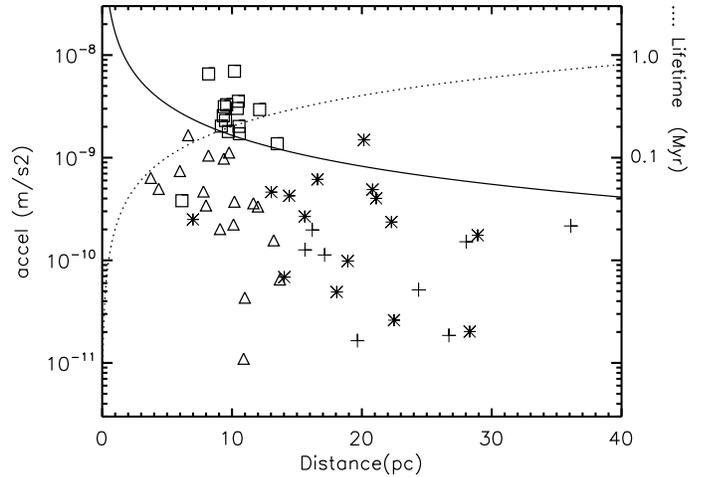}
\caption {Derived acceleration (in ${m}{s}^{-2}$) of clumps in the Rosette Nebula
  plotted against projected distance from the nebula centre (assumed to lie
  at $06^h 32^m 10.2^s +~04^o 49' 28''$, the location of the most
  luminous star HD46223). The solid line shows a model of acceleration
  by photoevaporation by the central O stars based on Bertoldi \&
  McKee (1990). The datapoints have been corrected for inclination
 assuming $i=30\degree$ (see text). The symbols
  represent the NW clumps (squares), other central clumps (triangles),
  and more distant compact clumps (stars). The crosses illustrate
the velocity gradients and apparent acceleration in the smooth
  background clouds, thought to be unaffected by the O~stars.}

}
\end{figure}

As well as the acceleration, the internal velocity dispersion 
($\Delta v$) is higher in the innermost clumps. For clumps
with $D_p \leq$13pc, $\Delta v = 1.6 \pm 0.44 km s^{-1}$,
marginally larger than the distant clumps ($1.2 \pm 0.46 km s^{-1}$).
It is not due to smearing of the velocity gradient due to the finite
beam size, which in
all cases would be less than $0.3 km s^{-1}$.
This suggests that the turbulence in the clumps is affected by
the interaction and acceleration of the O-star winds.
Line emission from the extended smooth clouds is narrower, with
widths of $1.0 \pm 0.4 km s^{-1}$.

We can estimate the clump physical conditions by examining the
$^{12}$CO 3-2/1-0 line ratios. However, the results of such analysis
are subject to limitations and uncertainties: the optical depths in these lines can be
large, the effects of PDRs can change the line
excitation, and the lines may be tracing gas at different depths in the clumps (with different kinetic temperatures, and/or resulting in different beam filling factors).  Molecular line ratios have been investigated in several similar regions, including detailed studies of individual clumps within the RMC (eg Patel et al., 1993;
Gonzalez-Alfonso \& Cernicharo, 1994; W95; White et al., 1997). PDR models have also been applied to a few individual lines of sight in the RMC on a 2~arcmin scale (see Schneider et al., 1998).  Although data from many transitions and
isotopomers are needed to uniquely interpret the results, the
large scale and dynamic range of the present data does allow us to investigate general
trends in the clumps across the whole of the GMC.

Using the J=1-0 $^{12}$CO data from Heyer et al. (2006), we include in
Table~2 the 3-2/1-0 line ratio of clumps and the clouds in the region, after
convolving the J=3-2 data to the same resolution (45~arcsec).  Note
that the objects chosen for this analysis are distinct isolated
features in the clouds, and are well separated from
the HV outflows. The clumps have ratios ranging from 0.4-1.8, and in
Figure~11, we plot the line ratio against the separation from the brightest
O~star HD46223 (roughly the luminosity centre of the nebula). In
general, this shows a decrease in ratio as a function of separation. The
distant, extended smooth clouds (shown by crosses)
have ratios significantly lower than the clumps.

\epsfverbosetrue
\epsfxsize=9.0 cm
\begin{figure}
\center{
\leavevmode
\epsfbox{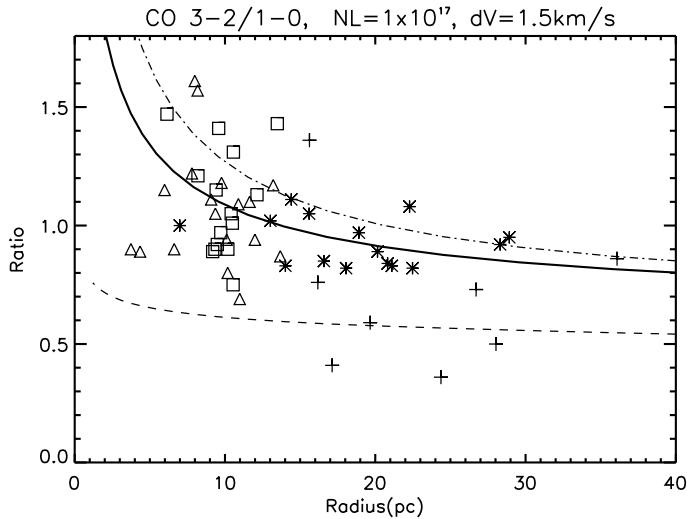}
\caption {Ratio of J=3-2 / 1-0 $^{12}$CO for clumps in the Rosette
  Nebula, plotted against the projected separation from the nebula
  and luminosity centre. The symbols
  represent the NW clumps (squares), other central clumps (triangles),
  more distant compact clumps (stars), and the centres of the extended
  distant clouds (crosses). The curves show the predicted line
  ratios using the radiative transfer model ``Radex'' for three
  different gas densities: $3\times10^3$ (dashed), $3\times10^4$
  (solid) and $3\times10^5 cm^{-3}$ (dot-dashed).  We use a constant CO
  column density of $10^{17} cm^{-2}$ and linewidth of $1.5 km s^{-1}$,
  and assume the gas and dust temperatures are equal.
  Kinetic temperature is derived from the separation from
  the luminous star (see text for details).}

}
\end{figure}

Using the program ''Radex'' (van der Tak et al., 2007) we can derive
3-2/1-0 line ratios for different gas densities and temperatures,
assuming an isothermal gas. With a CO column density
$N_L(CO)=10^{17} cm^{-2}$ and a linewidth of $1.5 km s^{-1}$,
this gives the line ratio diagram in Figure~12.  This
column density is
equivalent to an extinction, $A_V = 1.0$, assuming a CO abundance of
$10^{-4}$ and $N_L(H_2) / A_V =10^{21} cm^2 /magn$. 
$A_V = 1.0$ is considered a lower limit for most of the
clumps. In particular, studies of two of
the NW clumps show $A_V \sim 1-10$ (Gonzalez-Alfonso \&
Cernicharo, 1994; Patel et al., 1993). CO column densities higher
than $\sim 3\times 10^{17} cm^{-2}$
($A_V=3$) will result in high optical depths in both the 1-0 and 3-2
lines, and $^{12}$CO line ratios close to unity. That the observed 3-2/1-0
ratios in most clumps are higher than this suggests that most line
emission is from warm gas near the clump surface, at around $A_V \sim
1$, and without extreme line optical depths.
This suggests that the clumps are being heated
externally and CO is tracing the heated outer layer, similar
to the model for CO emission from cores near the
Orion cluster (Li et al., 2003).

\epsfverbosetrue
\epsfxsize=9.0 cm
\begin{figure}
\center{
\leavevmode
\epsfbox{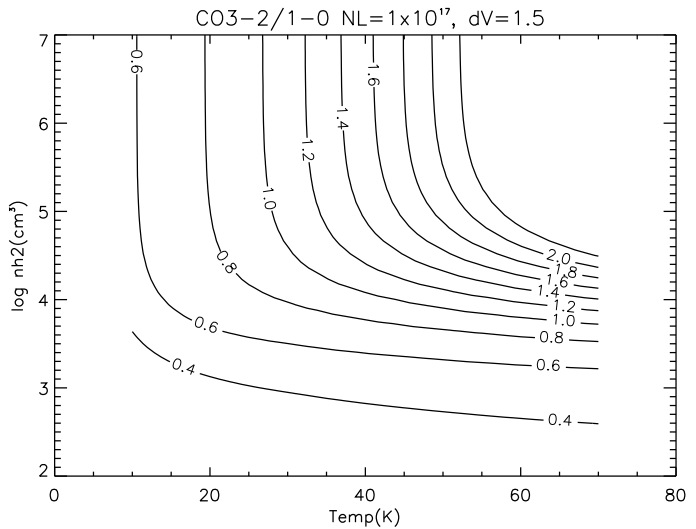}
\caption {Plot of J=3-2 / 1-0 $^{12}$CO line ratio as a function of
gas kinetic temperature and total volume density, using the ``Radex''
code (van der Tak et al., 2007). This uses a linewidth of $1.5 km
s^{-1}$ and CO column density of $10^{17} cm^{-2}$. For comparison,
most of the observed clumps have ratios of 0.5-1.6.}

}
\end{figure}

To compare the observed line ratio with the model, we include in Figure~11 three
curves showing the
predicted 3-2/1-0 ratio as a function of separation from the centre
of the nebula, assuming equal gas and dust temperatures and dust
heating by the central O~stars.
The luminosity centre of the nebula is assumed to be
at the location of the O4.9V star HD46223. In the interest of
simplicity, we assume that all clumps have similar mean volume
densities in
the emitting region, and ignore projection effects. We adopt a simple
power law for dust kinetic temperature in the clumps:
$$T = 35~K . [ L/{10^6L_{\odot}}]^{1/6} . [a/{0.1\mu m)}]^{-1/6} .
[R/{10pc}]^{-1/3}$$
This assumes heating by
central O stars of total luminosity $L = 10^6 L_{\odot}$ at a mean
location of the brightest star (HD46223), and a
grain emissivity based on Mie theory with a grain size $a = 0.1\mu m$
(eg Kr\"ugel, 2003). The predicted line ratio is plotted
as a function of $R$ for three values of the gas density, $n_{H_2}$. The
datapoints are consistent with typical gas densities of
$3\times10^4 cm^{-3}$ (shown by the solid line). This value is
sufficiently high that gas and dust will be closely coupled (eg Galli
et al., 2002), resulting in similar gas and dust temperatures.
Gas in the smooth distant clouds
(shown as crosses) lies below this line, and likely has a lower
density, or lies at a larger de-projected distance and is cooler than
predicted.

In summary, the plot shows that line ratios and derived clump
temperatures are consistent with radiative heating by the central
stars. As noted above, it is likely this is the clump surface
temperature; internal clump temperatures are not accessible to the
$^{12}$CO lines because of high optical depths, and their
determination would require measurements using less abundant isotopologues.

\subsubsection{The SE and S rims}

The sharply-delineated SE rim, known as the Monoceros Ridge,
is a clear example of the interface
between O~star radiation field and a Molecular Cloud (Blitz \& Thaddeus, 1980;
Schneider et al., 1998). As noted by Schneider et al., the region shows a sharp drop in molecular
emission on the side of the cloud facing the O stars. They also noted that it is composed of several discrete components in position-velocity space. In
Figure~13 we show a colour-coded image of the J=3-2 line in this region, as well
as the edge of the cloud to the S of the central cavity (the S rim).
Along both SE and S rims, the molecular cloud has a sharply-defined bright edge, spatially
unresolved on one side.
This indicates that the edge facing the O stars is $<$15~arcsec (or 0.1pc)
across, ie it is likely to be a thin structure viewed almost edge-on.  In the lower-resolution
J=1-0 images
(Heyer et al., 2006), the limb-brightening is less clear, and the
J=3-2/1-0 line ratio (after convolving the 3-2 data to the same
resolution) shows a peak at the bright edge of 1.1, dropping gradually into
the molecular cloud to $\sim 0.8$. This suggests that the limb brightening of
the J=3-2 line is due to a temperature peak at the cloud edge of 30K (averaged over
a 45~arcsec beam), dropping to $\sim$20K in the cloud core (cf Figure~12).

\epsfverbosetrue
\epsfxsize=8.0 cm
\begin{figure}
\center{
\leavevmode
\epsfbox{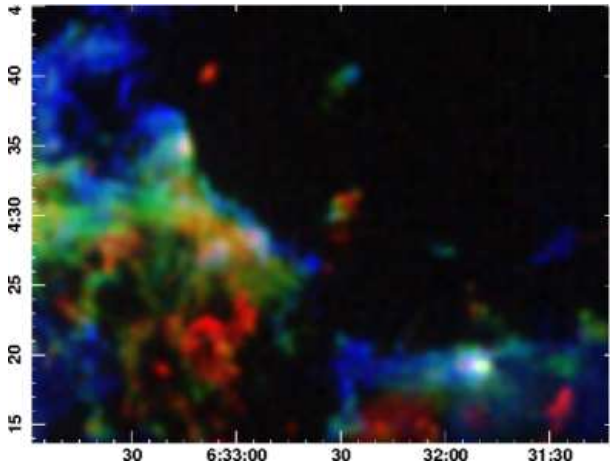}
\caption {Three-colour image of J=3-2 $^{12}$CO integrated intensity in the vicinity of
the SE and S rims, illustrating the velocity structure at the interface between the
molecular cloud and the central cavity.
The blue, green and red colours represent integrated intensities in
the velocity ranges 11.5-13.7, 13.7-15.8 and 15.8-17.9 $km s^{-1}$ respectively.
Resolution of this image is 20~arcsec, and the exciting O stars lie to the NW of
the map. The two bright broad-lined objects in the S and SE rims are associated with the outflows ROF4 and ROF7 and with IRAS objects.
}

}
\end{figure}

In addition to the sharp-edged structure,
the SE-S cloud rims show a velocity gradient, with
$\delta v/\delta r \sim 2 - 3 km s^{-1} pc^{-1}$. This is mostly orthogonal to the edge, and can be seen in the colour-coded image in Fig.~13, where the rim facing the O stars is blue-shifted, and the emission
becomes more red-shifted deeper into the RMC. The gradient is of similar
magnitude but opposite sign to that seen across the NW clumps (Table~2),
suggesting that the rims lie on the far side of the nebula, and that
gas is being accelerating away from us. This relative location along the line-of-sight
is confirmed by the absence of corresponding dark absorbing clouds
in optical images, and contrasts with
the NW side, where every blue-shifted CO
clump has a clear corresponding dark cloud (cf Figure~9).
The bulk molecular gas velocity of the
SE rim gas is $+15 km s^{-1}$, similar to that of the nebula and RMC
itself. Note that this is the projected velocity; the true velocity
may be higher if acceleration has occurred mostly in the plane of the
sky (for example, if it is being viewed edge-on).

In addition to a velocity shift, the datacube also shows
an increase in linewidth across the SE rim, from $1.2 km s^{-1}$ at the
blue-shifted edge, to $2.4 km s^{-1}$ towards the molecular cloud core. 
Although in some regions this broadening may be
due to multiple clumps at different velocities along the line of sight (eg Schneider et al., 1998), the general trend throughout the rims is that the clump gas velocities
are more dispersed further into the cloud.

Both the SE-S molecular rims and the NW ridge (see above)
are at a similar separation from the O stars, and are presumably experiencing a similar photoionisation flux. Other physical characteristics (CO linewidth, velocity gradient, temperature) 
are similar in both the SE-S molecular rims, and the NW clumps. 
However these regions clearly have different morphologies, with small
filling-factor clumps elongated in the radial direction in the NW (Fig.~9),
and a large filling-factor molecular rim orientated tangentially in
the SE-S (Fig.~13). The reason for this
difference is likely due to the different molecular gas densities in these two directions.
As the gas in the RMC to the SE disperses, the contiguous rim is likely to
evolve into a series of more dispersed isolated clumps, similar
to that currently found in the NW.

\subsubsection{Clump mass distribution}

The images of CO and optical emission (Figures~1, 7 \& 9)
show that much of the gas within 25~pc of the Rosette Nebula
appears highly clumped over a wide range
of size scales; several authors have also noted that
the RMC contains a large number of compact, low-mass clumps.
Molecular observations of the bright SE section of the RMC have been used to estimate
the clump mass distribution in this region (W95, Schneider et al., 1998).
However, the large area coverage and simultaneous
high spatial resolution of the present
dataset potentially allows us to measure the clump mass distribution over
a relatively large dynamic range, and to search for differences in the clump mass distribution over the cloud. To identify individual clumps, we have
employed the Starlink CUPID package (Berry et al., 2007),
with the 3-D Gaussian fitting algorithm, based on Stutzki \& G\"usten (1990).
This allows a direct comparison with similar analysis of other GMCs (eg Kramer et al., 1998, and refs. therein).
To minimise the ambiguity of clump identification and increase the robustness of the results, we set rather conservative
lower limits to the peak brightness temperature (5.5K, $T_{mb}$), clump size (20~arcsec), and linewidth ($0.7 km s^{-1}$). This technique identified 2050 clumps with these limits over the complete area shown in Fig.~1.

Reconstructing a model datacube with the identified clumps generates
$\sim 70\%$ of the {\em compact} emission flux ($<2$~arcmin across).
However, very little of the large-scale emission ($>5$~arcmin) from the
extended smooth clouds is picked up using this technique, and only 2$\%$ of the identified clumps have sizes larger than 5 beamwidths, as the Gaussian clump-fitting
method tends to subdivide larger clumps. Although it is
likely that the $^{12}$CO transition is optically thick - particularly
in the dense cores, we assume here that the integrated
intensity scales linearly with the mass (eg W95), and use
published $^{13}$CO studies of individual
clumps (where the masses can be measured more accurately) to scale the results.
In particular, Patel et al. (1993) have estimated the mass of
Globule~1 (listed as Clump~44 in Table~2), and Schneps et al. (1980) and
Gahm et al. (2006) measured the ``Wrench clump''
(Clump~4 in Table~2). Other isolated bright regions have
also been observed in $^{13}$CO in W95. For
those clumps with published masses, our integrated J=3-2 $^{12}$CO
intensity and the published masses scale linearly within a factor of 2
over the mass range 5-1000$M_{\odot}$. By adopting this scaling, we
derive the clump mass distribution for the whole mapped region shown
in the upper histogram of Figure~14.  This can be fit by a slope of
-0.8 (shown by the dashed line) for masses in
the range 3-100$M_{\odot}$. Although clumps (or ``Globulettes'') with
masses as small as $\sim 0.1 M_{\odot}$ can be detected in some sparse regions of the 
present datacube (see Fig.~9), the confusion limit is considerably higher in most regions, which results in the drop in the number counts below $\sim 3 M_{\odot}$ in Fig.~14.

The fitted slope in Fig.~14 implies a mass power law, $dN/dM \sim -1.8$, over the range
3-100$M_{\odot}$, for the whole RMC. This compares with -1.6 found by Schneider et al (1998) using two different datasets covering the SE region alone, and -1.3 measured by W95 for the SE region using a different clump-finding algorithm (although their technique was
noted by Kramer et al. to blend low-mass Gaussian clumps, resulting in a flatter slope). This value is similar to that found
for other large-scale molecular clouds (eg Kramer et al., 1998; Heyer
et al., 2001), suggesting there is no excess of compact, low-mass clumps in the Rosette Nebula, and that it is a typical GMC. There is marginal evidence of a steeper power law at the high mass end (above 100$M_{\odot}$), although this would need confirmation using a less abundant and hence optically thinner molecular species.

Figure~14 compares the mass distribution for clumps in the complete datacube
(within a projected separation of $\sim25$~pc from the O stars), clumps in the
inner region of the Rosette nebula (separation $<14$~pc),
and clumps in the innermost expanding ring (section 3.2). CO images (eg Fig.~9) show that these inner regions are apparently
highly-clumped, and are presumably more disrupted by the central O stars
(see section 3.3.1 above). However,
the three histograms show that not only is the power spectrum similar to that of other GMCs,
but there is no clear difference in the mass distribution of clumps close to the O stars
compared with that of the RMC as a whole. 

\epsfverbosetrue
\epsfxsize=9.0 cm
\begin{figure}
\center{
\leavevmode
\epsfbox{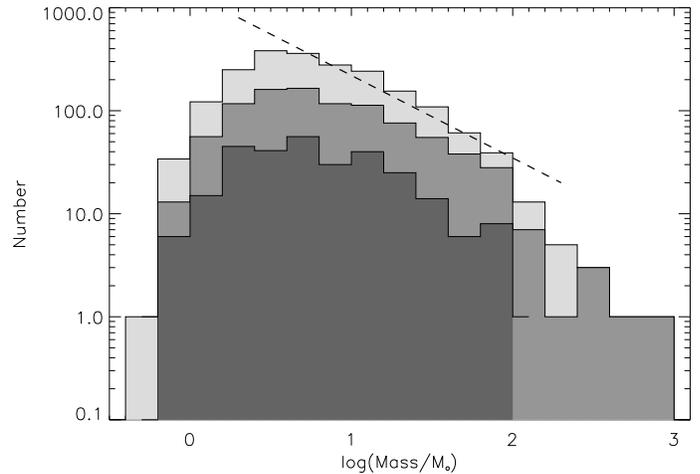}
\caption {Histograms of clump integrated intensity in the Rosette
  region. The upper faint grey plot includes clumps found over the whole map
  area, the middle
includes those for the inner region (within 14pc of
  the centre), and the lower darkest grey plot includes only the clumps in the
  expanding ring noted in section 3.2.
Masses are estimated by scaling from a published
  multi-isotopomeric study of one of the clumps (Patel et al.,
  1993). The dashed line represents a mass distribution $N(M) \propto
  M^{-0.8}$.  }

}
\end{figure}

\subsubsection{Association of molecular gas and young stars}

Seven of the 14 IRAS far-infrared objects
in the Rosette region were shown by W95 to be associated
with bright molecular gas clumps, and were presumed to be young
embedded luminous stars. The present CO study shows that 7 luminous
IRAS objects in the RMC also have associated
high-velocity outflowing gas (see Table~1), 6 of which
are common to the list of clumps in W95.

More recent infrared surveys have shown that several of these
IRAS objects may actually be composed of young stellar clusters rather
than individual luminous stars (eg Phelps \& Lada, 1997; Poulton et al.,
2008).  Moreover, as noted in section~3.3.2 above, the CO clumps
identified in W95 decompose into many sub-clumps at
increased spatial resolution.  Poulton et
al. (2008) and Balog et al. (2007) conducted a detailed
3~--~70$\mu m$ continuum study of a significant fraction of the Rosette region
with the Spitzer telescope. By comparing their data with our
CO maps, we can examine the association between the young infrared
objects and the molecular gas at increased resolution and towards
lower luminosity stars. This is best done at 24$\mu m$, where the
resolution is high enough to identify individual stars, and the
youngest, most embedded objects can be seen.

Figure~15 shows a superposition of the infrared-excess stars identified
by Poulton et al. on the J=3-2 CO peak intensity.  The
crosses show stars with SEDs fitted by Poulton et al. using
a circumstellar disk model (which we loosely term SED Class II);
the squares represent those which
require an additional envelope to fit the excess (here called Class
I).  Many (but not all) of the Class I (envelope) objects are associated with
clumps of
molecular gas. There are also clearly some regions -- such as that around the
main cluster NGC2244 itself, in the NW of this image -
which contain many Class II (disk)
objects, but which are devoid of CO. Conversely, there are regions with
apparently bright and clumpy molecular
gas but which have very few Class I or II stars; examples are the
clumps 15-40~arcmin SSE
of the main cluster.  Both Poulton et al. and Balog et al.
noted the dearth of Class I
but abundance of Class II objects within the cluster NGC2244.
Can these different sets of observations be reconciled?

CO emission from a disk-only (Class II) source would not be detected
with the present observations: for a typical disk of 300~AU radius,
an optically thick line and a mean temperature of 40K,
beam dilution would give $T_{mb}\sim
0.04K$, an order of magnitude below the current observational limit.
However, envelopes around young stars typically have radii an order of
magnitude larger, and therefore would be detectable in our data.
Looking at their individual
Class I objects, 30/36 in the mapped area have nearby CO
emission (within 30~arcsec).  However, this is formally an upper limit
to the association of gas and young stars, as in some cases
the CO may just lie along the line-of sight to the star.

Using the same $I_{CO}$:$M_{clump}$ scaling factor as
measured above (see section
3.3.3), we can estimate gas mass limits
around the 6 Class I objects identified
by Poulton et al. using complete SEDs, which have {\em no}
associated CO. The resultant upper limits for their clump masses are $\leq
0.2M_{\odot}$.  However, it may be that
these objects are misclassified, and are actually Class II disks viewed
edge-on; a highly-inclined disk can mimic the SED of
envelope-dominated structure due to the high extinction to the central
star. In this case, we would not expect to detect CO.
The relatively small number of Class I objects without associated
CO (6, compared with 750 Class II objects) is
consistent with this intepretation, as long as the disks are
geometrically thin.

\epsfverbosetrue
\epsfxsize=9.0 cm
\begin{figure}
\center{
\leavevmode
\epsfbox{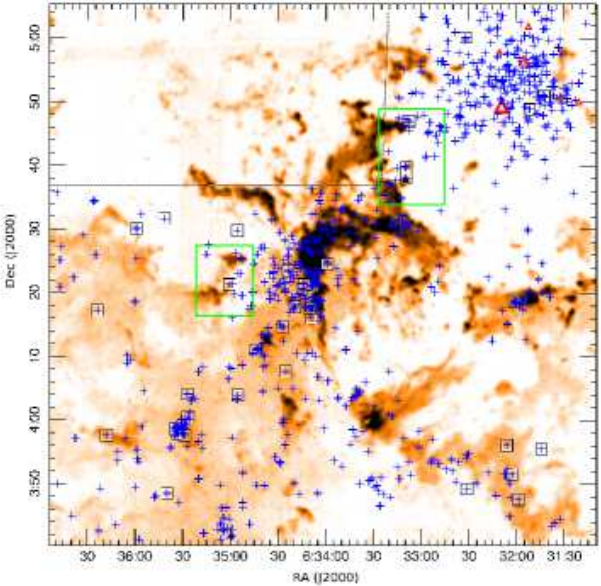}
\caption {Map of peak J=3-2 $^{12}CO$ emission, with the
  IR-excess stars identified by Poulton et al. (2008)
  superimposed. The dashed line towards the upper left marks the boundary of the area studied
  in the infrared. The blue crosses represent stars with an infrared
  excess, where the SED
  can be fit by a disk, and those shown by squares require
  an addition envelope component. The large red triangles represent
  the O stars illuminating the Rosette Nebula. The two green rectangles
  are the areas illustrated in more detail in the middle and right panels of
  Figure~16.}  }
\end{figure}

Poulton et al. required detections at multiple wavelengths to classify
the objects. However, several additional compact objects can be seen in their
longer-wavelength data, which were not formally identified as
YSOs, either due to incomplete sky coverage at all
wavelengths or because they were too faint at some wavelengths.
Examples can be seen in the NW ridge, where compact 24-${\mu}m$ objects
are associated with the tips of the 2 or 3 of the CO clumps (see
Fig.~16, left panel). Optical
images of the tip of the Wrench Clump also show evidence of an
embedded star or compact reflection nebula (Gahm et al., 2006).  Many of the
associations of YSOs with CO clumps are found near the bright rims
at the edges of the molecular
clouds, and resemble the young stars found in some
evaporating globules within the Eagle Nebula (eg McCaughrean \&
Andersen, 2002). Figure~16 shows in detail some of the clearer examples of CO
clumps (given by the contours) and their association with 24$\mu m$
point sources and bright-rimmed clouds. As argued
above, these peaks in the CO emission are
too bright to be arise from a compact
disk, and are likely to be evaporating envelope gas. This is
confirmed by their IR classification as envelope-dominated in several
cases (see Poulton et al.).

If the $24\mu$m compact sources in the NW globules are also confirmed
as young stars, it is likely they will have a similar velocity as that
of the CO gas -- typically, the relative velocity of stars and their
gas envelope is $< 1 km s^{-1}$ (Covey et al., 2006).  This would
imply that the NW ridge is forming stars which have a relative velocity
of $15 km s^{-1}$ compared with the rest of the RMC. This compares
with the escape velocity of the RMC of $\sim 6 km s^{-1}$, implying
this will eventually eject a ring of young stars.

\epsfverbosetrue
\epsfxsize=19.0 cm
\begin{figure*}
\center{
\leavevmode
\epsfbox{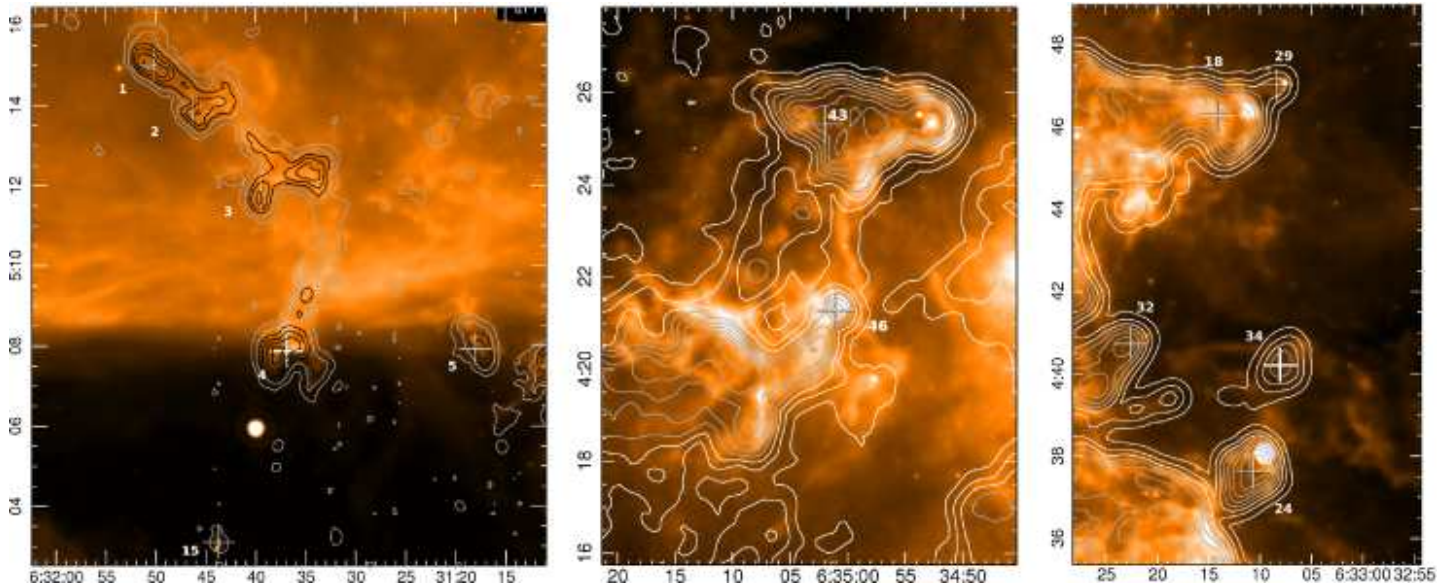}
\caption {Superposition of the Spitzer $24 \mu m$ image
on the peak J=3-2 CO emission (contours) in some of the
regions with Class I and II stars identified by Poulton et al (2008)
or where 24$\mu m$ point-like sources can be identified, and which
have associated clumps in the CO datacube. The crosses show some
of the CO clumps and their identifications in the list in
Table~2. CO contours start at 2.7K, with intervals of 2.7K ($T_{mb}$).
Axes are RA/Dec (J2000).}
}
\end{figure*}

\section{Discussion - the effects of the OB stars on the molecular cloud}

The molecular clouds in the vicinity of the Rosette Nebula, as
revealed in the new CO images, show a highly clumpy morphology, with
sharp-edged clouds mostly facing the OB stars. The overall clumpy structure
is very similar to the simulated images from SPH modelling of the effects of
O-star photoionisation on a turbulent molecular cloud (cf Dale
et al., 2007, Fig.~2). Their models suggested that the
effect of the luminous stars on the surrounding star formation rate
is rather modest; triggered star formation in such circumstances is
not a very significant phenomenon.

The results of the present large-scale mapping of the molecular gas
shows a good correlation between luminous young
clusters (with luminosity $\geq 10^3 L_{\odot}$), high velocity outflows
and bright regions of molecular emission.  Moreover in several cases,
individual stars with envelopes are also associated
with bright CO clumps. In most cases, these are at the edges or rims
of the GMC, facing the central nebula.
A particularly clear example is Clump 44 in
Table~2, seen near the top of the middle panel in Figure~16.
Here a point-like 24-$\mu m$ source lies at the centre of the head of the CO
clump. Patel et al. (1993) identified an IRAS object near this location,
with a luminosity of $\sim 100L_{\odot}$ (although this may be an
upper limit as the $100 \mu m$ flux is confused). Patel et al. and
White et al. (1997) noted that there was a significant
velocity gradient along the length of this globule.

The new data (Figure~17) shows that this velocity gradient is remarkably
constant, at $1.7 km s^{-1} pc^{-1}$
over a distance of 1.7pc (270~arcsec) behind the star. It is unclear at present
whether the RDI model (eg Bertoldi \& McKee, 1990; Lefloch \& Lazareff,
1994) can predict such an extended and constant acceleration in the molecular
gas behind such clumps; in general most of their results show
curvature in the position-velocity diagrams.
Furthermore the RDI model predicts significant
velocity gradients only in the relatively short-lived
(and hence rare) ``pre-cometary phase''. The present results (see
Table~2) show that 
significant velocity gradients are the norm rather than the exception
for exposed molecular clumps.

\epsfverbosetrue
\epsfxsize=9.0 cm
\begin{figure}
\center{
\leavevmode
\epsfbox{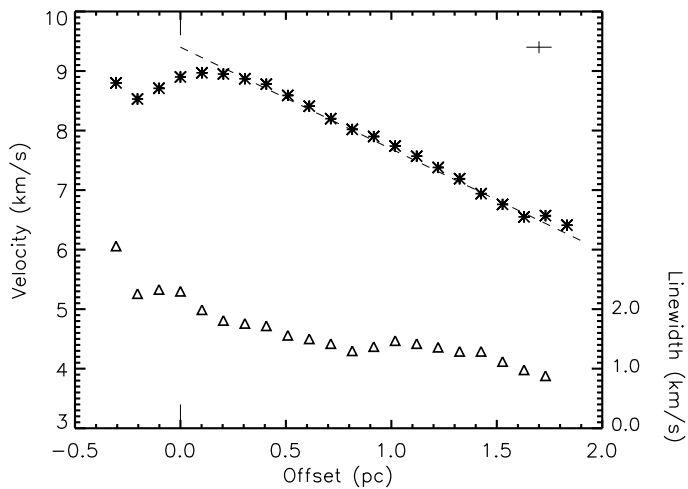}
\caption {Velocity (shown by star symbols) and width of the CO line
  (FWHM, shown by triangle symbols) along the extended Clump 44. The
  values are taken by fitting a Gaussian
  to spectra extracted from the datacube every
  15~arcsec along the length of the clump. The abscissa is the offset at PA
  70$^o$, relative to the bright 24$\mu m$ source (shown by the
  vertical bars). A typical error bar is shown in the upper right. The
  dashed line shows a constant velocity gradient of $1.72 km s^{-1} pc^{-1}$.
}
}
\end{figure}

A bright 24$\mu m$ source lies near the centre of the head of
Clump 44 and, in general,
the clearest associations of young stars and CO clumps are found near
the cloud edges (see Figure~16 for other examples).
What fraction of the CO clumps have embedded young stars? This depends
on the infrared sensitivity and hence the stellar luminosity limit of
the observations and is, moreover, relatively insensitive to the most
highly-embedded objects.
But if we use the 24-$\mu$m Spitzer data to trace young
stars and -- in the absence of longer-wavelength, high-resolution data
-- assume that the luminosity scales with the $24 \mu m$ flux, then we
can estimate the luminosity detection limit in these data by scaling the
bolometric luminosity of the Clump~44 star. From this, we estimate a
luminosity limit of $\sim 10 L_{\odot}$ - equivalent to a $1.5
M_{\odot}$ (or F-type) star at 2~Myr age. In Table~2 we have indicated
the clumps where a
compact 24$\mu m$ source has been identified within the CO emission.
Approximately 40\% of these clumps have such an object; Figure~18
shows that CO clumps with stars (indicated by the
filled histogram) tend to have somewhat higher
velocity gradients compared with clumps without stars. This 
suggests that velocity gradients - which we associate with higher
photoionisation and clump acceleration - are
also related to the presence of a young, relatively luminous
star. Although the data cannot detect
lower-mass stars, it does suggest that the formation of early-type
stars (spectral class F or earlier) and gas acceleration are linked.
More sensitive observations of the {\em apparently} starless clumps
would be of interest to determine whether all molecular clumps with
velocity gradients contain
stars, and whether the stellar mass is related to the local gas acceleration.

\epsfverbosetrue
\epsfxsize=8.0 cm
\begin{figure}
\center{
\leavevmode
\epsfbox{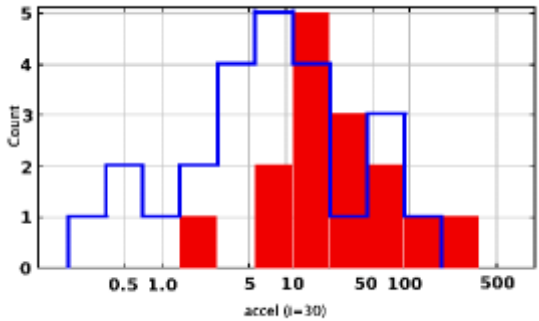}
\caption {Histogram showing the distribution of radial
 CO clump acceleration
 (in $km s^{-1} pc^{-1}$) for clumps with stars (red
  shaded histogram) and clumps without stars (open
  histogram). This assumes the same inclination from the plane of the
 sky for all clumps ($i=30\degree$).
 Clumps with associated stars tend to have the higher apparent acceleration.}
}
\end{figure}

Does the presence of a CO clump and envelope associated with the
embedded 
stars necessarily imply a more massive molecular clump was originally
present, and that these are the youngest objects in the region?
The photoevaporation lifetime of a clump
depends on its radius, $r_c^{-1.5}$,
so compact, denser clumps will survive longer (and
will be accelerated more gradually, in the RDI model)
than larger, diffuse clumps. 
Cores containing stars are thought to have a more centrally-peaked
structure than starless cores (eg Ward-Thompson et al., 1994), so for
the same clump mass, we might expect clumps with a central star
would be more centrally-peaked and hence
survive longer in the UV environment, compared with a more diffuse starless
core. In such circumstances, we would naturally expect to see that
many of the longer-lasting clumps will be the ones containing
young stars. Rather than the
stars still associated with molecular clumps and envelopes
being the youngest objects, it would suggest that
the presence of the star causes the molecular clump to survive longer in the UV
field. Further observations would be of interest to look for other differences
between the clumps with and without young stars.

\section{Conclusions}

A large-scale single-dish imaging survey covering 60$\times$60~pc of the Rosette Nebula and Rosette Molecular Cloud has been carried out with 14~arcsec resolution in the J=3-2 transition of $^{12}$CO.
The resulting datacube has a high spatial dynamic range (the ratio of map size to spatial
resolution is $\sim500$), and reveals the complex and clumpy spatio-velocity
structure in the molecular gas throughout the RMC.
The following conclusions can be drawn:

\begin{itemize}

\item{We identify $\sim$2000 clumps with a mass distribution $dN/dM
  \propto M^{-1.8}$, independent of distance from the central O stars.
This is similar to other less disrupted clouds,
  and suggests the clump mass distribution is not strongly affected by the
  O stars.}
\item{Comparison of the molecular datacube with optical and infrared
images show that all the gas which blue-shifted with respect to the cloud
systemic velocity lies in the foreground. Red-shifted gas has no dark
absorbing foreground counterpart.
The dominant gas motion is therefore expansion from the cluster of O stars.}
\item{The locations and velocities of many of the clumps closest to
  the O stars can be fit by an 11pc radius expanding ring.
This has a dynamical timescale of $\sim 1$~Myr, similar to the
  nebula age. Stars formed in clumps in this ring are not dynamically
  bound to the system.}
\item{Most molecular clumps show significant radial velocity gradients,
  generally radial from the O stars. Velocity gradients
tend to be larger for clumps closer to the O stars, and can be explained by
acceleration
  through photodissociation of the cloud edges facing the central
  stars. However, in one of the clearest cases of an extended clump,
the velocity
  gradient is constant over a distance of $\sim$1.7pc, and it is unclear
  whether this
acceleration mechanism can produce such a constant velocity gradient. }
\item{Comparison with 24-$\mu$m Spitzer images shows many examples of
  embedded young stars within the accelerating clumps. These are
  examples of evaporating envelopes, and we find an association
  between the high velocity gradients and the presence of an embedded
  star.}
\item{The CO 3-2/1-0 ratio in the clumped gas decreases with distance from
  the O stars, implying a radial decrease in temperature.
Most of the emission from the
  $^{12}$CO line is thought to arise from gas on the surface of the
externally-heated clumps, with the main energy source being
the central O stars.}

\end{itemize}

\section*{Acknowledgments}

The James Clerk Maxwell Telescope is operated by the Joint Astronomy
Centre on behalf of the United Kingdom Particle Physics and Astronomy
Research Council, the Netherlands Organisation for Scientific
Research, and the National Research Council of Canada. We wish to
acknowledge help from the many people involved in the construction of
ACSIS and HARP, and the development of the upgrades to the JCMT
control system.

We also wish to thank Nick Wright for making available his H$\alpha$
mosaic of the Rosette Nebula.

\section{References}

\parindent 0mm

Alvarez, C., Hoare, M., Glindemann, A., Richichi, A., 2004, \aa, 427, 505

Aspin, C., 1998, \aa, 335, 1040

Bachiller, R., Fuente, A., Kumar, M.S.N., 2002, \aa, 381, 168

Balog, Z., Muzerolle, J., Rieke, G.H., Su, K.Y.L., Young, E.T.,
Megeath, S.T., 2007, \apj, 660, 1532

Berry, D.S., Reinhold, K., Jenness, T., Econonou, F., 2007, in
Astronomical Data Analysis Software and Systems XVI, ed. R.A. Shaw,
F. Hill \& D.J. Bell, ASP Conference Series 376, 425

Bertoldi, F., McKee, C.F., 1990, \apj, 354, 529

Beuther, H., Schilke, P. Sridharan, T.K., Menten, K.M, Walmsley, C.M.,
Wyrowski, F., 2002, A\&A 383, 892

Blitz, L., Thaddeus, P., 1980, \apj, 241, 676

Blitz, L., Stark, A.A., 1986, \apj, 300, L89

Borkin, M., Arce, H., Goodman, A., Halle, M. 2008, in Astronomical Data Analysis Software and Systems XVII, ed. R.W. Argyle,
P.S. Bunclark \& J.R. Lewis, ASP Conference Series 394, 145

Celnik, W.E., 1985, A\&A 144, 171

Covey, K.R., Greene, T.P., Doppmann, G.W., Lada, C.J., 2006, \aj, 131, 512

Dale, J.E., Clark, P.C., Bonnell, I.A., 2007, MNRAS, 377, 535

Drew, J. et al., 2005, MNRAS, 362, 753

Gahm, G.F., Carlqvist, P., Johansson, L.E.B, Mikolic, S., 2006, A \&
A, 454, 201

Gahm, G.F., Grenman, T., Fredriksson, S., Kristen, H., 2007, \aj, 133, 1795

Galli, D., Walmsley, M., Gon\c calves, J., 2002, \aa, 394, 275

Gonzalez-Alfonso, E., Cernicharo, J., 1994, \apj, 430, L125

Gonzalez-Alfonso, E., Cernicharo, J., Radford, S.J.E., 1995, \aa, 293, 493

Hensberge, H., Pavlovski, K., Verschueren, W., 2000, \aa, 358, 553

Herbig, G.H., 1974, PASP, 86, 604

Hester, J.J. et al., 1996, \aj, 111, 2349

Hester, J.J. \& Desch, S.J., 2005, ASP conf. proc. 341, ``Chondrites and
the Protoplanetary Disk'', ed. Krot, A.N., Scott, E.R.D., Reipurth, B., 107

Heyer, M.H., Carpenter, J.M., Snell, R.L., 2001, \apj, 551, 852

Heyer, M.H., Williams, J.P., Brunt, C.M., 2006, ApJ, 643, 956

Hills, R.E., et al., (in prep.)

Hovey, G.J., Burgess, T.A., Casorso, R.V., Dent, W.R.F., Dewdney,
P.E., Force, B., Lightfoot, J.F., Willis, A.G., Yeung, K.K., 2000,
SPIE, 4015, 114

Jenness, T., Cavanah, B., Economou, F., Berry, D.S., 2008, in
Astronomical Data Analysis Software and Systems XXX, ed. J. Lewis,
R. Argyle, P. Bunclark, D. Evans, \& E. Gonazle-Solares,
ASP Conference Series (in press)

Kramer, C., Stutzki, J., R\"ohring, R., Corneliussen, U., 1998, \aa,
329, 249 

Kr\"ugel, E., 2003, ``The Physics of Interstellar Dust'', IOP
Publishing, Bristol

Lada, C.J., Gautier, T.N., 1982, \apj, 261, L161

Lefloch, B., Lazareff, B., 1994, \aa, 289, 559

Lefloch, B., Lazareff, B., 1995, \aa, 301, 522

Li, D., Goldsmith, P.F., Menten, K., 2003, \apj, 587, 262

Li, J.Z. \& Smith, M., 2005, \apj, 130, 721

Li, J.Z., Smith, M.D., Gredel, R., Davis, C.J., Rector, T.A., 2008, \apjl, 679, L101

MacLow, M., Klesse, R.S., 2004, Review Modern Physics, 76, 125

McCaughrean, M.J., Andersen, M., 2002, \aa, 389, 513

McKee, C.F., 1989, \apj, 345, 782

McKee, C.F., Ostriker, E.C., 2007, Ann. Rev. Astr. Astrophys., 45, 565

Oort, J.H. \& Spitzer, L., 1995, \apj, 121, 6

Patel, N.A., Xie, T, Goldsmith, P.F., 1993, \apj, 413, 593

Patel, N.A., Goldsmith, P.F., Snell, R.L., Hezel, T., Xie, T., 1995,
\apj, 447, 721

Phelps, R.L., Lada, E.A., 1997, \apj, 477, 176

Phelps, R.L., Ybarra, J.E., 2005, \apj, 627, 845

Poulton, C.J., Robitaille, T.P., Greaves, J.S, Bonnell, I.A.,
Williams, J.P., Heyer, M.H., 2008, \mnras, 384, 1249

Pound, M.W., 1998 \apj, 493, L113

Rees, N. et al., 2002, SPIE, 4848, 283

Schneider, N., Stutzki, J., Winnewisser, G., Blitz, L., 1996, \apjl, 468, L119

Schneider, N., Stutzki, J., Winnewisser, G., Block, D., 1998, \aa,
335, 1049

Schneps, M.H., Ho, P.T.P., Barrett, A.H., 1980, ApJ, 240, 84

Smith, H. et al., 2003, SPIE, 4855, 338

Stutzki, J., G\"usten, R., 1990, \apj 356, 513

van der Tak, F.F.S., Black, J.H., Schoir, F.L., Jansen, D.J., van
Dishoeck, E.F., \aa, 468, 627

Ward-Thompson, D., Scott, P.F., Hills, R.E., Andre, P., 1994, MNRAS,
268, 276

White, G.J., Lefloch, B., Fridlund, C.V.M., Aspin, C.A., Dahmen, G.,
Minchin, N.R., Huldtgren, M., 1997, \aa, 323, 931

Williams, J.P., Blitz, L, Stark, A.A., 1995, ApJ 451, 252 (W95)

Wu, Y., Wei, Y., Zhao, M., Shi, Y., Yu, W., Qin, S., Huang, M., 2004,
A\&A 426, 503

Ybarra, J.E., Phelps, R.L., 2004, \aj, 127, 3444

\label{lastpage}

\end{document}